\documentclass[11pt,twoside,a4paper,cmspaper,final,collab]{cms-tdr}
\def\svnVersion{1419cb6-D}\def\svnDate{2026/09/17}\def\cmsCernNoTag{CERN-EP-2026-185}\def\cmsCernDate{\today}\def\cmsMessage{Submitted to the Journal of High Energy Physics}
\begin{document}\cmsNoteHeader{EXO-24-039}

\providecommand{\Pj}{\HepParticle{j}{}{}\xspace}
\newcommand{\mjj}{\ensuremath{\overline{m}_{2\Pj}}\xspace}
\newcommand{\mjjjj}{\ensuremath{m_{\mathrm{4j}}}\xspace}
\newcommand{\DatasetLumi}{\ensuremath{138\fbinv}\xspace}
\newcommand{\atrue}{\ensuremath{\alpha_{\text{true}}}\xspace}
\newcommand{\deepjet}{\ensuremath{\textsc{DeepJet}}\xspace}
\newcommand{\PSuu}{\HepParticle{S}{\PQu\PQu}{}\xspace}
\newcommand{\PStwothree}{{\HepParticle{S}{2/3}{}}\xspace}
\newcommand{\PY}{\HepParticle{Y}{}{}\xspace}
\newcommand{\PSQtst}{\HepSusyParticle{\PQt}{}{*}\xspace}

\cmsNoteHeader{EXO-24-039}

\title{Search for resonant and nonresonant production of pairs of dijet resonances with \texorpdfstring{\PQb}{b} jets in the final state in proton-proton collisions at \texorpdfstring{$\sqrt{s} = 13\TeV$}{sqrt(s) = 13 TeV}}

\date{\today}

\abstract{
   A search is presented for pair-produced dijet resonances of identical mass, each decaying into a bottom (\PQb) quark and a light-flavor quark. The search uses a data sample corresponding to an integrated luminosity of \DatasetLumi collected with the CMS detector in proton-proton collisions at $\sqrt{s}=13\TeV$. Results are presented separately for nonresonant production and for the case where the four-jet production proceeds via an intermediate resonant state.  Upper limits are reported on the production of four-jet and dijet resonances with \PQb jets in the final state. This is the first LHC search for resonant production of pairs of fully resolved dijet resonances with \PQb jets in the final state, in which all resonances considered correspond to non-standard-model states. For nonresonant production, the most significant excess occurs at an average dijet mass of 1.0\TeV, with a local significance of 2.1 standard deviations. Additionally, the results exclude pair production of top squarks with masses between 0.5 and 0.8\TeV, extending previous searches for $R$-parity-violating decays to strange and \PQb quarks. In the search for resonant production, the most significant excess occurs at a four-jet resonance mass of 3.1\TeV and a dijet resonance mass of 1.3\TeV, with a local significance of 2.6 standard deviations. The results are also used to set limits at 95\% confidence level on the diquark-to-diquarks model and exclude the heavier diquark masses between 2 and 7\TeV. 
}

\hypersetup{
pdfauthor={CMS Collaboration},
pdftitle={Search for resonant and nonresonant production of pairs of dijet resonances with b jets in the final state in proton-proton collisions at sqrt(s) = 13 TeV},
pdfsubject={CMS},
pdfkeywords={CMS, paired dijets, resonances, b jets}}

\maketitle

\section{Introduction}
\label{sec:IntroAndMotivation}

Many theories of physics beyond the standard model (BSM) predict the existence of a new particle, \PX, that decays into pairs of jets~\cite{Hill:1991at,Randall:1999ee,Kribs:2007ac,Kilic:2008pm,Robertpaper,Duminica:2025lte}. The simplest searches conducted at the CERN LHC are for $s$-channel single production of dijet resonances, and place tight constraints on many models of BSM physics~\cite{ATLAS:2015nsi,dijet13TeV,Sirunyan:2016iap,Aaboud:2017yvp,Sirunyan:2018xlo,Aad:2019hjw,Sirunyan:2019vgj}. A natural extension involves a more complex process with pair production of dijet resonances yielding four resolved jets in the final state.  These searches for nonresonant and resonant production of pairs of dijet resonances in the high-mass region have only recently begun at the LHC and thus present a promising opportunity for discovery~\cite{DijetPaper,PairedDijetATLAS}. Here we search for pair production of dijet resonances whose decays contain heavy flavors, specifically bottom quarks (\PQb) that hadronize to form \PQb jets ($b$) in the final state.

In this analysis, we perform searches for pairs of dijet resonances that have a narrow natural width and include \PQb jets in the final state, using a data sample corresponding to an integrated luminosity of \DatasetLumi collected with the CMS detector in proton-proton ($\Pp\Pp$) collisions at a center-of-mass energy of 13\TeV. Searches for nonresonant production of pairs of dijet resonances with final states that include \PQb jets have been previously performed by the CMS and ATLAS Collaborations, both using around 36\fbinv of data \cite{AlePaper,ATLAS:2017ofn}.  We present the first such search for resonant production, and we extend the previous nonresonant search using the data collected by CMS in 2016--2018. Additional motivation for this search comes from the previous CMS search of this type without flavor tagging~\cite{DijetPaper}, which yielded an excess with  a local (global) significance of 3.9 (1.6) standard deviations (s.d.) for a signal hypothesis where a resonance of mass 8.6\TeV decays to a pair of dijet resonances of mass 2.15\TeV each. Moreover, that CMS search also observed an excess with a local (global) significance of 3.6 (2.5) s.d.~for a signal hypothesis of nonresonant production of pairs of dijet resonances with a mass of 0.95\TeV each. An extension of the previous untagged search~\cite{DijetPaper}, targeting broad four-jet resonances, yielded a local (global) significance of 3.9 (2.2) s.d. for a signal hypothesis in which a four-jet resonance with a mass of 3.6\TeV and a relative width of $\Gamma/M = 10\%$ decays into two dijet resonances, each with a mass of 1\TeV. Thus, this analysis also serves to investigate these excesses within the context of heavy-flavor scenarios, in which each dijet system includes a \PQb jet.

\begin{figure}[htbp!]
\centering
\includegraphics[width=0.49\textwidth]{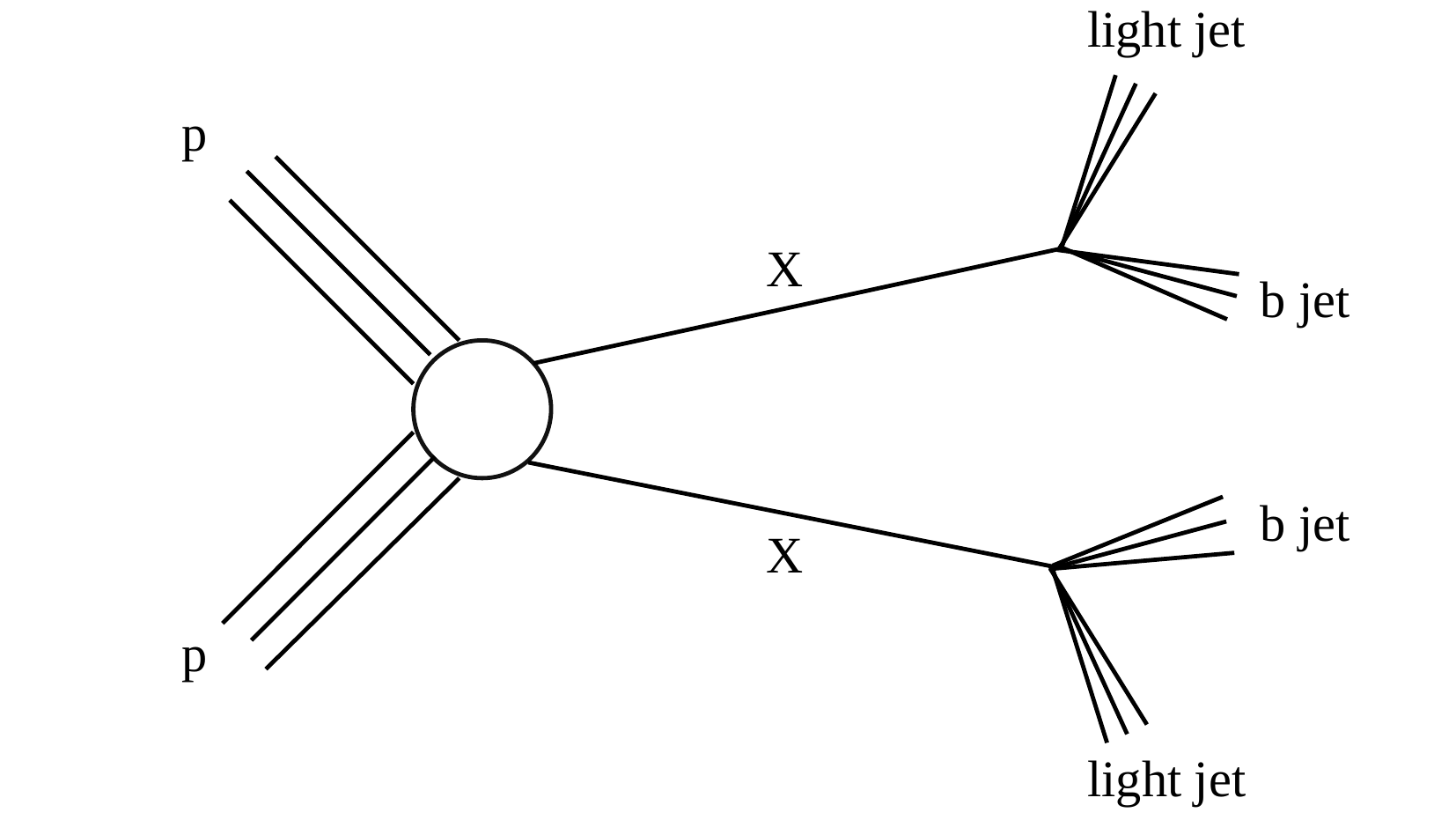} 
\includegraphics[width=0.49\textwidth]{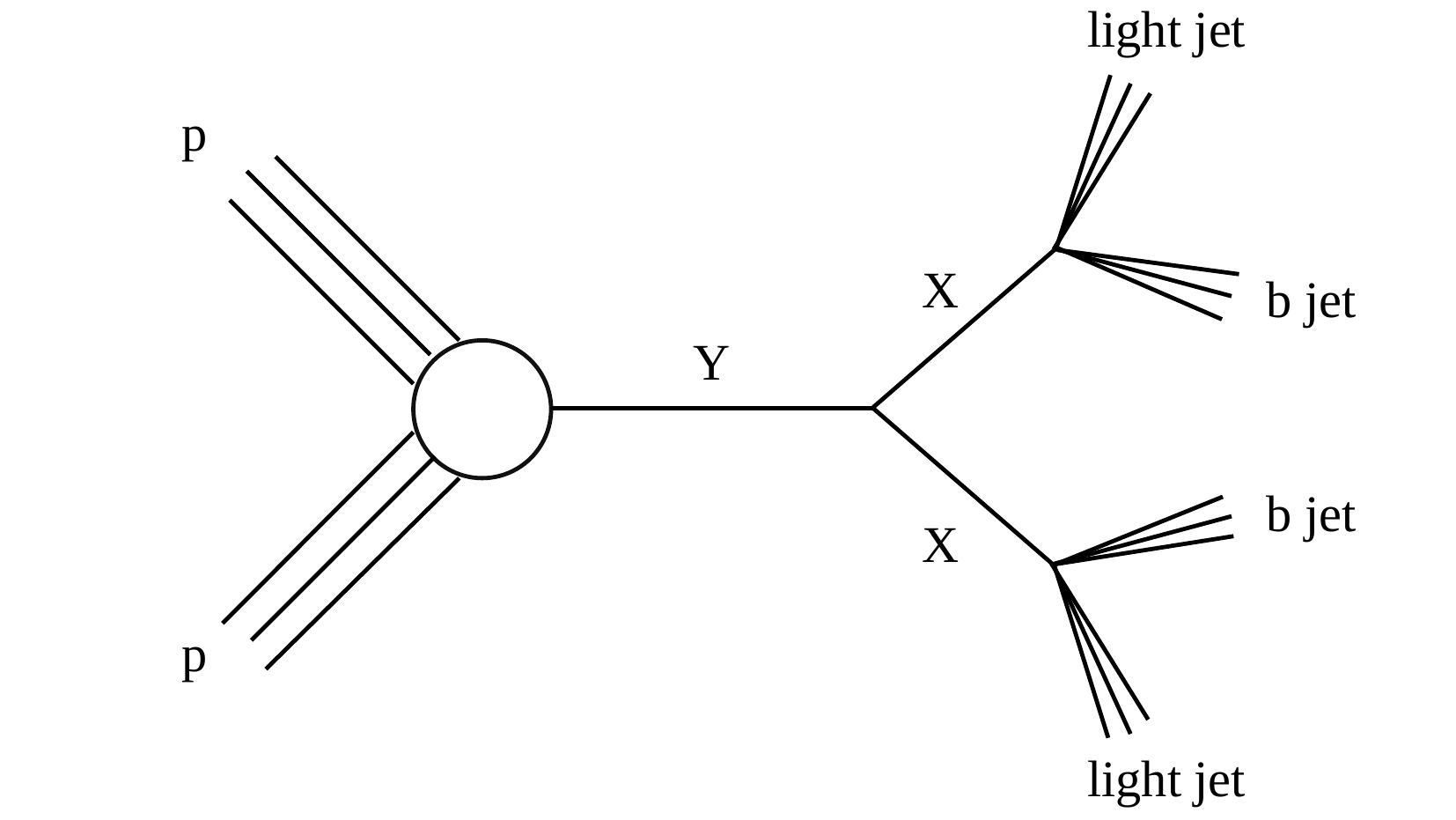}
\caption{(Left) Nonresonant production of pairs of dijet resonances, \PX. (Right) Resonant production of pairs of dijet resonances, \PX, via a massive resonance, \PY. In both scenarios, each dijet resonance, \PX, produces a \PQb jet and a light-flavor jet.}
\label{fig:feynman}
\end{figure}

We perform a search for two different physics scenarios, each with a distinct production mode, both depicted in Fig.~\ref{fig:feynman}. The first scenario is the nonresonant pair production of resolved dijet resonances with spin-0 particles, \PX, each yielding a \PQb jet and a light-flavor (up, down, strange quark) jet (\Pj), 
\begin{equation}
  \Pp\Pp \to \PX\PX \to (\PQb \Pj)(\PQb \Pj).
  \label{eq:nonres}
\end{equation}
The second scenario we consider is the resonant pair production of dijet resonances, \PX, via a massive BSM particle, \PY, where again each \PX produces a \PQb jet and a light-flavor jet:
\begin{equation}
  \Pp\Pp \to \PY \to \PX\PX \to (\PQb \Pj)(\PQb \Pj).
  \label{eq:res}
\end{equation}
We consider the $R$-parity-violating (RPV) supersymmetry model~\cite{Evans2012,Barbier:2004ez} as a benchmark for the nonresonant search, where dijet resonances, the top squark (\PSQt) and its antisquark (\PSQtst), are pair-produced. Each of the pair-produced top squarks decays to a  strange (\PQs) and bottom quark via the trilinear coupling with strength $\lambda''_{UDD}$, between the up-type squark and the two down-type quarks. Specifically, the coupling strength is $\lambda''_{323}$, where the numerical subscript $323$ refers to the quark generations of \PSQt, \PQs, and \PQb quarks: 
\begin{equation}
    \Pp\Pp \to \PSQt\PSQtst \to (\PAQb\PAQs)(\PQb\PQs).
\end{equation}
For the resonant search, we consider the diquark-to-diquarks (D2D) model~\cite{Bogdanpaper} as a benchmark, where a heavy color-sextet scalar diquark resonance (\PSuu) of electric charge 4/3, produced from the fusion of two up quarks, decays to two identical color-triplet scalar diquarks (\PStwothree) of electric charge 2/3, and each of the \PStwothree decays to a \PAQb and \PAQs quark:
\begin{equation}
    \Pp\Pp \to \PSuu \to \PStwothree\PStwothree \to (\PAQb\PAQs)(\PAQb\PAQs).
\end{equation}
This analysis explores the pair production of dijets, each containing a \PQb jet, with the same data set and a similar analysis strategy as the previous search without flavor tagging~\cite{DijetPaper}. The search is based on the four-jet invariant mass and average dijet mass distributions, where a signal would appear as a localized enhancement over the expected background. 

The paper is organized as follows. Section 2 gives a brief description of the CMS detector, and Section 3 describes the simulated signal and background samples used in the analysis. The event reconstruction and selection criteria are presented in Section 4, followed by the analysis strategy in Section 5. The modeling of the signal and the background estimation are described in Sections 6 and 7, respectively, and the associated systematic uncertainties are discussed in Section 8. Finally, the results are presented in Section 9, followed by the summary of the paper in Section 10. Tabulated results are provided in the HEPData record for this analysis~\cite{HEPData}.

\section{The CMS detector}
\label{sec:CMSDetector}

The CMS apparatus~\cite{CMS:2008xjf,CMS:2023gfb} is a multipurpose, nearly hermetic detector, designed to trigger on and identify electrons, muons, photons~\cite{CMS:2020cmk,CMS:2016ngn,CMS:2024aqx}, and (charged and neutral) hadrons~\cite{CMS:2020uim,CMS:2018rym,CMS:2014pgm}. A global ``particle-flow'' (PF) algorithm~\cite{CMS:2017yfk} aims to reconstruct all individual particles in an event, combining information provided by the all-silicon inner tracker and by the crystal electromagnetic, and the brass and scintillator hadron calorimeters, operating inside a 3.8\unit{T} superconducting solenoid, with data from the gas-ionization muon detectors embedded in the flux-return yoke outside the solenoid. The reconstructed particles are used to build \PGt leptons, jets, and missing transverse momentum~\cite{CMS:2018jrd,CMS:2016lmd,CMS:2019ctu}.

Events of interest are selected using a two-tiered trigger system. The first level (L1), composed of custom hardware processors, uses information from the calorimeters and muon detectors to select events at a rate of around 100\unit{kHz} within a fixed latency of 4\mus~\cite{CMS:2020cmk}. The second level, known as the high-level trigger (HLT), consists of a farm of processors running a version of the full event reconstruction software optimized for fast processing, and reduces the event rate to a few kHz before data storage~\cite{CMS:2016ngn,CMS:2024aqx}. 

\section{Simulated data samples}
\label{sec:SimulatedDataSamples}

The dominant background in both the resonant and nonresonant searches originates from quantum chromodynamics (QCD) multijet production. Simulated QCD multijet samples are used only to optimize event selections that maximize the expected signal significance, and are not used in the background estimation itself, which is performed based exclusively on data, as described in Section~\ref{sec:BackgroundEstimation}. Contributions from other standard model processes, such as top quark pair (\ttbar), single top quark, and electroweak multiboson production, are expected to be small after the event selection and are included implicitly in the background estimate. Simulated data samples of QCD multijet background are only used in order to design optimized event selections that maximize the expected signal significance. The QCD samples are produced at leading order (LO) with $2 \to 2$ jet processes, with additional jets from initial- and final-state radiation included in the parton shower. This sample is produced using \PYTHIA~8.205 with the CUETP8M1 underlying event tune~\cite{Skands:2014pea,Khachatryan:2015pea}, and NNPDF2.3LO~\cite{Ball:2012cx} is used as the parton distribution function (PDF) set.

For the resonant search, the benchmark model is based on the D2D model.  Specifically, we consider the production of a scalar diquark resonance (\PSuu) that decays to two color-triplet scalars (\PStwothree), each decaying into a \PQb quark and a light quark. The branching fraction of this decay is set to 100\%. The \MGvATNLO~2.6.5~\cite{Alwall2014} generator is used to produce these events at LO, and \textsc{MadSpin}~\cite{Artoisenet:2012st} is used for the \PStwothree decay process. Samples are produced for multiple values of the diquark masses within the intervals $2 < M(\PSuu) < 8\TeV$ and $0.11 < M(\PStwothree)/M(\PSuu) < 0.42$. 

The coupling of the \PSuu to pairs of up quarks is set to 0.4, the same value as used in the previous search~\cite{DijetPaper}. For the couplings of the \PSuu to pairs of \PStwothree, we use values determined for each \PSuu and \PStwothree mass combination, such that the \PSuu width stays equal to 0.43\% of its mass, the same value of the width as in the previous search where the \PSuu decays to pairs of vector-like quarks~\cite{DijetPaper}.

For the nonresonant search, the benchmark model used is the pair production of hadronically decaying RPV top squarks. We model signal at LO using \MGvATNLO~2.6.0~\cite{Alwall2014}, and events are generated with up to one additional parton in the matrix element calculations with the MLM jet matching scheme~\cite{mlmmatch}. The top squark decay occurs via the RPV interaction with coupling $\lambda''_{323}$, where each top squark decays to \PQs and \PQb quarks, as described in Section~\ref{sec:IntroAndMotivation}, and the branching fraction of this decay is set to 100\%: $\mathcal{B}(\PQb\PQs)=1$.  All superpartners, except the top squarks, are taken to be decoupled, and no intermediate particles are produced in the top squark decay. The couplings considered are large enough to ensure that the resulting decays occur promptly. We produce samples with top squark masses between 0.5 and 3\TeV.  

Both benchmark models use \PYTHIA~8.205~\cite{Sjostrand:2014zea} for parton showering, with the CP5 underlying event tune~\cite{m_2020_extraction} using the NNPDF3.1 PDF set~\cite{Ball:2014uwa} at next-to-next-to-LO precision in perturbative QCD. For all samples, the simulation of the CMS detector is done by \GEANTfour~\cite{geant4}, and effects from pileup, including additional pp interactions from the same or adjacent bunch crossings, are added.  The simulated pileup distribution is weighted to match that observed in data.

\section{Event reconstruction and selection}
\label{sec:EventReconstruction}

Particles are identified by the PF algorithm as muons, electrons, photons, charged or neutral hadrons. The jets are reconstructed, where the input particles are clustered using the anti-\kt algorithm~\cite{Cacciari:2008gp, Cacciari:2011ma} with distance parameters of 0.4 and 0.8. The primary vertex (PV) is taken to be the vertex corresponding to the hardest scattering in the event, evaluated using tracking information alone, as described in Section 9.4.1 of Ref.~\cite{CMS-TDR-15-02}. Charged PF candidates not originating from the PV are removed prior to the jet finding.

Events with loose jet requirements satisfying the single-jet L1 triggers are passed to the HLT system, which requires the events to exceed a minimum \HT threshold. Here, \HT is defined as the scalar sum of the jet \pt for all jets reconstructed with a distance parameter of 0.4 with $\pt > 30\GeV$, $\abs{\eta} < 3.0$. For data collected in 2016 (2017--2018), the HLT requires events to have $\HT > 800\,(1050)\GeV$ or contain at least one jet with a distance parameter of 0.8 and jet $\pt > 450\,(550)\GeV$. The offline analysis is conducted using only reconstructed jets clustered with a distance parameter of 0.4. 

Jet momenta are determined as the vectorial sum of all particle momenta in the jet. Tracks identified as originating from pileup vertices are discarded and an offset correction is applied to correct for remaining contributions~\cite{CMS:2020ebo}. Jet energy corrections are derived from simulation studies such that the average measured energy of jets becomes identical to that of particle-level jets. In-situ measurements are used to determine any residual differences between the jet energy scale in data and in simulation, and appropriate corrections are made~\cite{CMS:2016lmd}. Additional selection criteria are applied to each jet to remove jets potentially dominated by instrumental effects or reconstruction failures~\cite{CMS:2020ebo}. 

The four jets with the highest \pt are defined as the leading jets in an event. Selected events are required to have all four leading jets with $\pt > 80\GeV$ and $\abs{\eta} < 2.4\,(2.5)$ for data collected in 2016 (2017--2018). The four leading jets are additionally required to pass identification criteria~\cite{CMS-PAS-JME-16-003} to veto jets associated with misreconstructed or isolated muons or electrons, and veto spurious jets arising from calorimeter noise. To pass these identification criteria, jets are required to have less than 90\% of the total energy coming from neutral hadrons and photons, less than 80\% of the total energy coming from muons, less than 80\% of the total energy from electrons, and a nonzero contribution to the total energy from charged hadrons.

The four leading jets are paired together to form two dijet systems, subsequently referred to as dijet 1 and dijet 2. A jet pairing algorithm is used to select the optimal pairing by minimizing spatial separation between the jets. This algorithm uses the $\eta\text{-}\phi$ separation, $\Delta R_{1}$ and  $\Delta R_{2}$, between the jets in each of the dijet systems, defined as:
\begin{equation}
  \Delta R_{1} = \sqrt{(\eta_{\mathrm{\Pj_a}} - \eta_{\mathrm{\Pj_b}})^2 + (\phi_{\mathrm{\Pj_a}} - \phi_{\mathrm{\Pj_b}})^2},\quad
  \Delta R_{2} = \sqrt{(\eta_{\mathrm{\Pj_c}} - \eta_{\mathrm{\Pj_d}})^2 + (\phi_{\mathrm{\Pj_c}} - \phi_{\mathrm{\Pj_d}})^2},
  \label{eq:deltaR1}
\end{equation}
where the subindices $\mathrm{\Pj_a}$ and $\mathrm{\Pj_b}$ refer to jets in dijet 1, and the subindices $\mathrm{\Pj_c}$ and $\mathrm{\Pj_d}$ refer to jets in dijet 2. There are three possible ways to pair four jets into two dijet systems, and this jet pairing algorithm selects the jet pairing configuration which minimizes the variable, $\Delta R$, where $\Delta R$ is defined as:
\begin{equation}
  \Delta R = \abs{\Delta R_{1} - 0.8} + \abs{\Delta R_{2} - 0.8}.
  \label{eq:deltaR_def}            
\end{equation}
The minimization of $\Delta R$ in Eq.~(\ref{eq:deltaR_def}) is based on the expectation that jets originating from the same dijet resonance are more likely to be close together than uncorrelated jets. The algorithm minimizes the separation between jets, and since the jets are reconstructed with a distance parameter of 0.4, the offset of 0.8 in Eq.~(\ref{eq:deltaR_def}) helps to suppress pairings where jets overlap in the $\eta\text{-}\phi$ plane. The efficiency of pairing jets correctly depends on the dijet resonance mass and whether the two dijets are produced resonantly or not, as discussed in detail in Section~\ref{sec:SignalSim}.

Once the optimal pairing is obtained, the following selections are made. Requirements on the separation of the jets within each
of the two dijets, $\Delta R_{1} < 2.0$ and $\Delta R_{2} < 2.0$, are imposed to reject the multijet QCD background. To further suppress $t$-channel QCD background, compared to the $s$-channel signal, the $\eta$ separation between the two dijet systems is required to satisfy:
\begin{equation}
  \Delta \eta = \abs{\eta_{1} - \eta_{2}} < 1.1,
  \label{eq:deltaEta}
\end{equation}
where $\eta_{1}$ and $\eta_{2}$ are the pseudorapidities of the first and second dijet
systems. Since masses of the resonances decaying to dijets are identical, we require the mass asymmetry of the two dijet systems to be small:
\begin{equation}
  \text{asymmetry} = \frac{\abs{m_{1} - m_{2}}}{m_{1} + m_{2}} < 0.1,
  \label{eq:massAsym}
\end{equation}        
where $m_{1}$ and $m_{2}$ are the masses of the two dijet systems. These selection criteria were optimized to maximize the expected signal significance, using Monte Carlo simulations of signals and QCD multijet production. The signal acceptance after imposing these requirements is discussed in Section~\ref{sec:SignalSim}.

The \PQb jets are identified using \PQb tagging algorithms~\cite{BTV-16-002}. The jets originating from \PQs quarks in our benchmark models cannot be distinguished from jets originating from \PQu or \PQd type quarks, and are defined as light-flavored jets. This analysis employs the \deepjet~\cite{Bols:2020bkb} \PQb tagger, which has an improvement of 5--20\% in the \PQb tagging efficiency for the same misidentification probability for light quarks compared to the \PQb tagger used in the earlier paired dijet analysis with \PQb jet final states~\cite{AlePaper}. Different \deepjet discriminator scores are considered for tagging the \PQb jets, and the number of \PQb jets required in the final state are varied to maximize the expected signal significance across all resonance masses. From these studies, we determine that the best expected signal significance is obtained by requiring each of the dijets in the event to have at least one \PQb jet. Here a jet is classified as a \PQb jet if its \deepjet score exceeds the `medium' \PQb tagging working point. This working point corresponds to approximately 75\% \PQb tagging efficiency using jets with $\abs{\eta} < 2.5$ and $\pt > 90\GeV$ in simulated \ttbar events, and approximately 1\% (20\%) misidentification probability for a light-flavored quark or gluon (charm quark) jet~\cite{CMS-DP-2023-005}. The efficiency to tag both of the \PQb jets decreases as the signal resonance mass increases since the resulting \PQb jets have higher \pt at higher masses, ranging from 38\% to 24\% for RPV top squark masses in the range 0.5 to 2\TeV, and from 52\% to 20\% for \PSuu masses in the range 2 to 8\TeV and \PStwothree masses in the range 0.5 to 2\TeV.

\section{Analysis strategy}
\label{sec:AnalysisStrategy}

\begin{figure}[htbp!]
\centering
\includegraphics[width=0.9\textwidth]{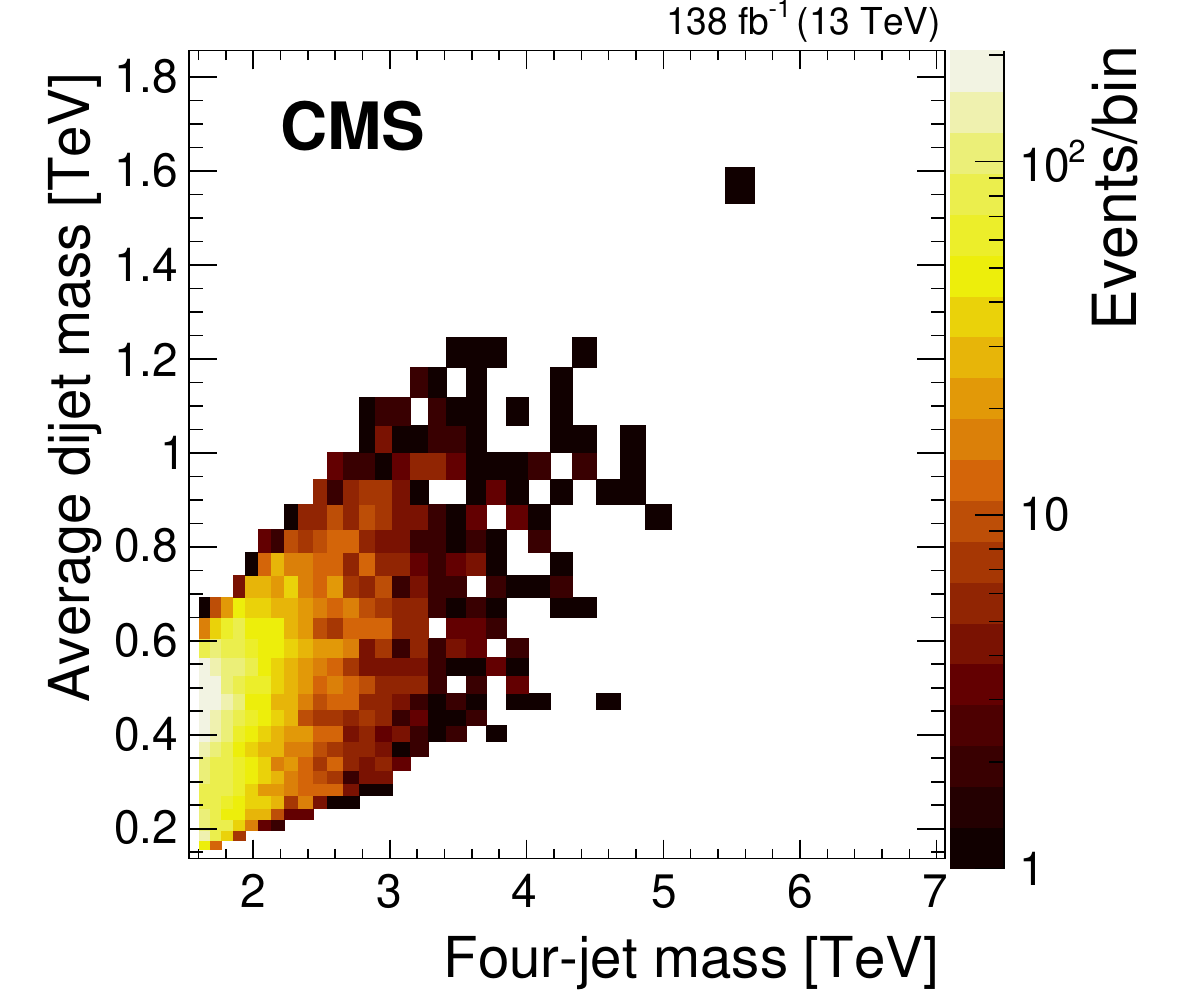} 
\caption{Number of events observed (color scale) within bins of the invariant mass of the four jets and the average dijet mass.}
\label{fig:2DPlot4JMvs2JM}
\end{figure}

This analysis considers the production of two identical particles, \PX, either via resonant or nonresonant mechanisms, as shown by Eqs.~(\ref{eq:nonres}) and (\ref{eq:res}). For the nonresonant case, there is a localized excess in the average dijet mass distribution, $\mjj = (m_{1} + m_{2})/2$. For the resonant signal, $\PY \to \PX\PX$, there is a localized excess in both the four-jet mass, \mjjjj, corresponding to the mass of \PY, and the  average dijet mass, \mjj, corresponding to the mass of \PX. Figure~\ref{fig:2DPlot4JMvs2JM} shows all the observed events in this plane, which pass the nominal kinematic selections and the \PQb tagging requirements described in Section~\ref{sec:EventReconstruction}. We observe the highest mass event at $\mjjjj = 5.5\TeV$ and $\mjj = 1.54\TeV$, further discussed in Section~\ref{sec:Results}. The two events with \mjjjj values near 8\TeV that predominantly contributed to the excess observed at a resonance mass of 8.6\TeV in the previous search without flavor tagging~\cite{DijetPaper} are not selected in this analysis with the \PQb tagging requirements applied. As the sensitivity of this analysis decreases at higher resonance masses, the search is not extended beyond resonance masses of 8\TeV and therefore cannot confirm or exclude the previously observed excess at 8.6\TeV. 

The dominant SM background contribution in these searches comes from QCD multijet production, but due to large theoretical uncertainties in the perturbative QCD predictions, the background is instead estimated directly through data. For this, the observed \mjj spectra in the nonresonant search and \mjjjj spectra in the resonant search are parameterized as steeply falling, smooth distributions, and the searches are performed by looking for signal-like localized excesses in these distributions.  The SM background contribution from \ttbar production is found to be $<$10\% of the total background, and the shapes in both the \mjj and \mjjjj spectra are smoothly falling, and is thus accounted for by the background estimation method based on fitting the data.

\begin{figure}[htbp!]
\centering
\includegraphics[width=0.49\textwidth]{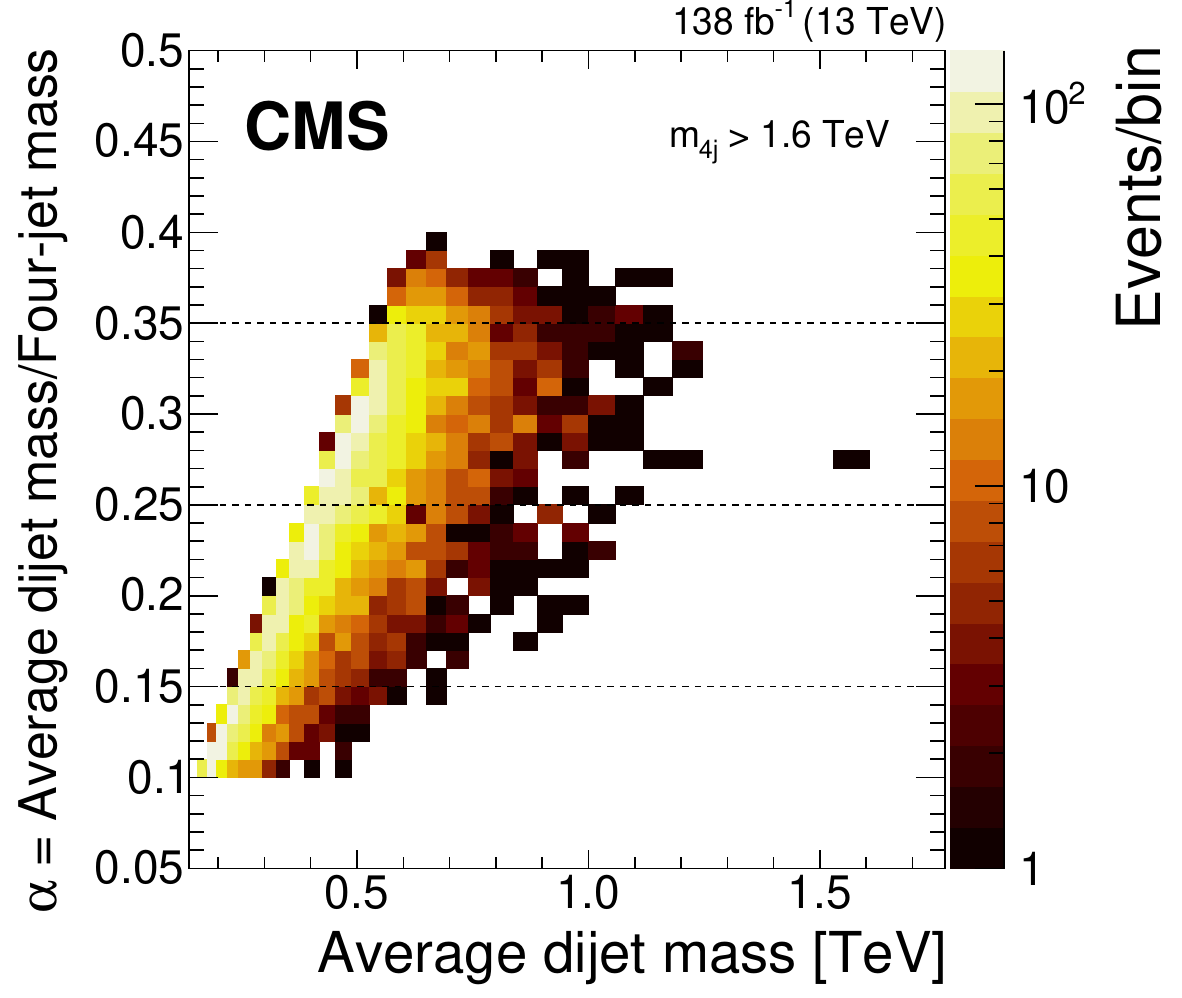} 
\includegraphics[width=0.49\textwidth]{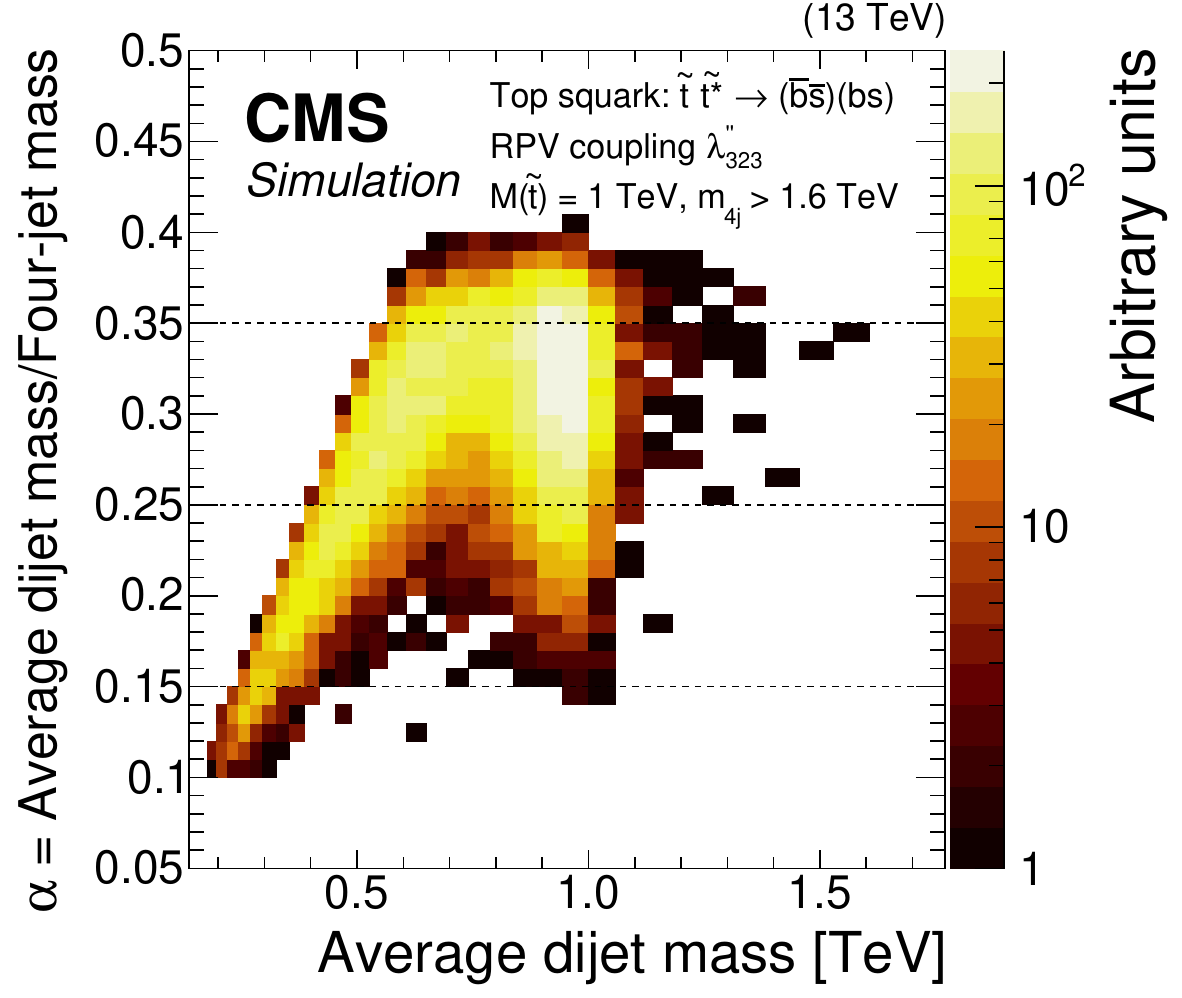}
\caption{(Left) Number of events observed (color scale) within bins of the average dijet mass and the ratio $\alpha$. (Right) Number of events predicted in the same bins by a simulation of the production and RPV decay of a pair of top squarks with a mass of 1\TeV. The distribution of the simulated signal events arises from the performance of the jet pairing algorithm, and is explained in Section~\ref{sec:SignalSim}. The dashed lines indicate the lower edges of the three $\alpha$ bins in which the analysis is conducted. 
}
\label{fig:2DPlotAlphavs2JM}
\end{figure}

\begin{figure}[htbp!]
\centering
\includegraphics[width=0.49\textwidth]{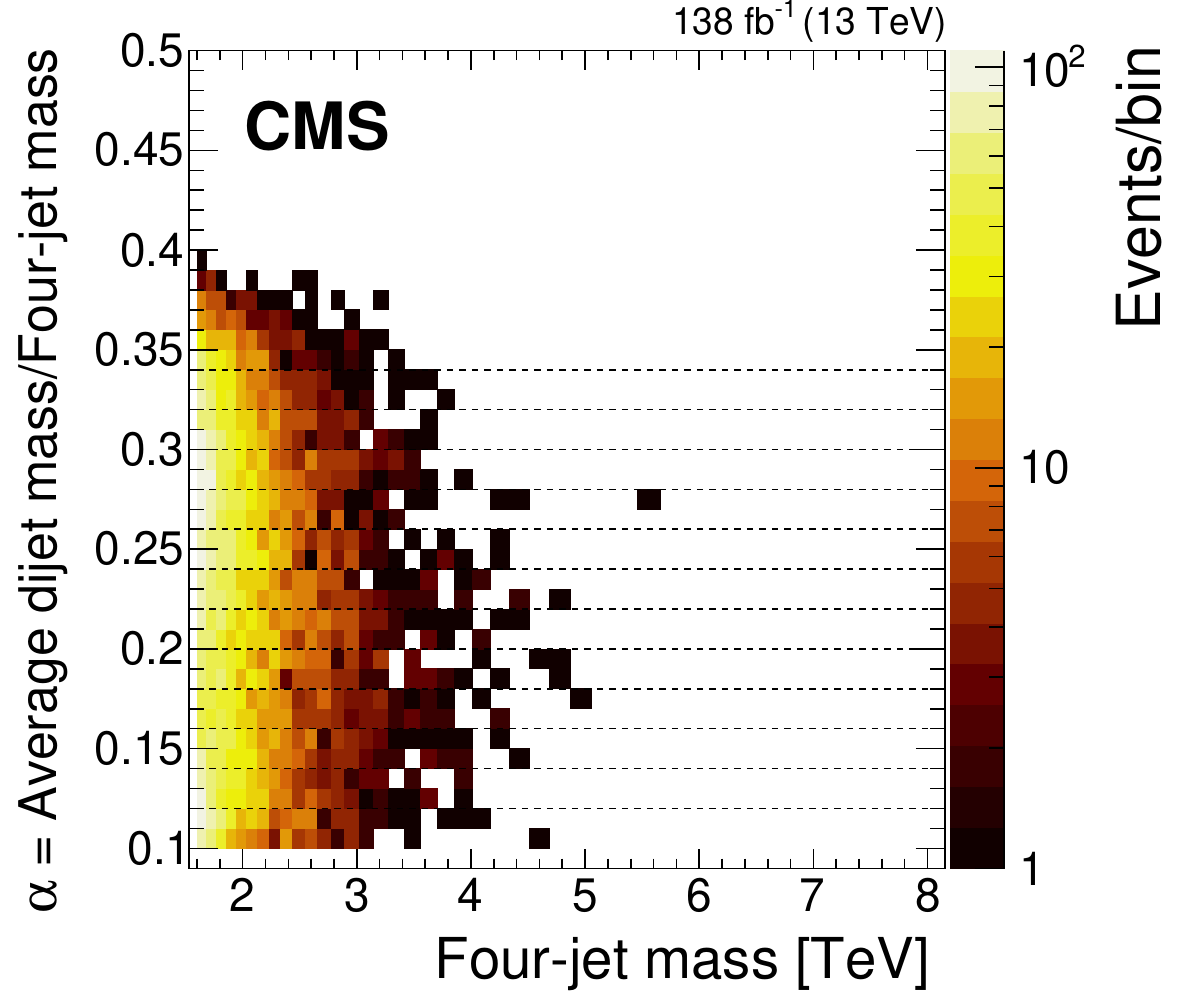} 
\includegraphics[width=0.49\textwidth]{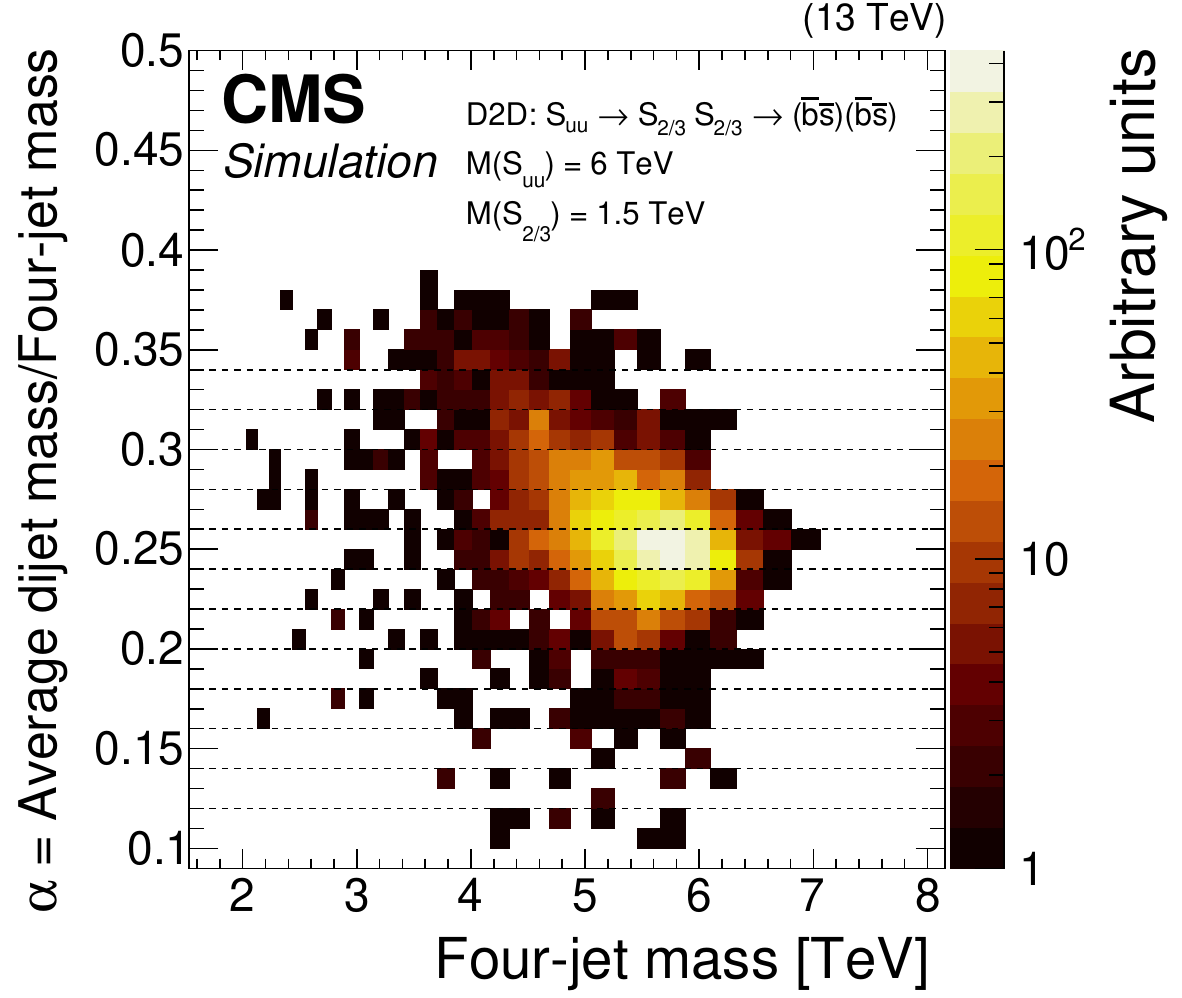}
\caption{(Left) Number of events observed (color scale) within bins of the four-jet mass and the ratio $\alpha$. (Right) Number of events predicted in the same bins by a simulation of a diquark with a mass of 6\TeV, decaying to a pair of color-triplet scalar diquarks, each with a mass of 1.5\TeV. The dashed lines indicate the lower edges of the twelve $\alpha$ bins in which the analysis is conducted.}
\label{fig:2DPlotAlphavs4JM}
\end{figure}

The distributions of \mjj for the nonresonant search and \mjjjj for the resonant search are binned in a dimensionless variable, $\alpha = \mjj / \mjjjj$. This is done because \mjj and \mjjjj are correlated variables, and binning \mjj into bins of \mjjjj or vice versa causes sculpting of these distributions, as observed in Fig.~\ref{fig:2DPlot4JMvs2JM}, making the accessible mass range and turn-on behavior highly bin-dependent. Figures~\ref{fig:2DPlotAlphavs2JM} and \ref{fig:2DPlotAlphavs4JM} show the two-dimensional distributions of $\alpha$ vs. \mjj for the nonresonant case, and $\alpha$ vs. \mjjjj for the resonant case, for both the observed data and signal simulation. While Fig.~\ref{fig:2DPlot4JMvs2JM} clearly shows the aforementioned correlations in \mjj vs. \mjjjj, Fig.~\ref{fig:2DPlotAlphavs4JM} demonstrates the decorrelation due to the usage of the $\alpha$ variable. 

Although a nonresonant signal is not expected to have a fixed $\alpha$ value, binning the data in $\alpha$ results in an enhanced signal sensitivity by exploiting the expected $\alpha$ distribution in signal. Properly reconstructed signal events, where the reconstructed average dijet mass agrees with the generated top squark mass, occur more frequently at higher $\alpha$ values, as seen in Fig.~\ref{fig:2DPlotAlphavs2JM} (right). This is especially true for heavier top squarks, as events with low values of $\alpha$ only occur when the top squarks are produced with high \pt.

For the nonresonant production scenario, we analyze three distributions of \mjj, corresponding to three bins with lower $\alpha$ edges of 0.15, 0.25, and 0.35, respectively. The number of bins and bin widths are determined through optimization studies that yield a width of 0.1 for the two bins of $\alpha$ between 0.15 and 0.35, and an overflow bin is utilized for $\alpha > 0.35$. A search for a mass bump, indicative of a dijet resonance, is conducted in each of these distributions simultaneously, where a single parameter of interest is considered across all bins, corresponding to the signal cross section. In each $\alpha$ bin, the analysis is performed only for \mjj values above a threshold for which the triggers are found to exceed 99\% efficiency. These mass thresholds are 0.35, 0.53, and 0.61\TeV for the lower $\alpha$ edges of 0.15, 0.25, and 0.35, respectively.

For the resonant production scenario, we examine twelve distributions of \mjjjj, with lower $\alpha$ edges ranging from 0.12 to 0.34, binned in intervals of 0.02, which is comparable to the reconstructed resolution of $\alpha$. An overflow bin is utilized for all $\alpha > 0.34$, and $\alpha$ values less than 0.12 are not included due to the large background contribution from \ttbar production. A search for a mass bump indicative of a potential four-jet resonance is conducted in the twelve distributions simultaneously, where a single parameter is considered for the signal cross section across all fits. Binning the data by intervals of $\alpha$ maximizes the signal significance and allows for a broader and less biased range of \mjjjj. The analysis is performed in the region $\mjjjj >1.6\TeV$, where the triggers are found to be greater than 99\% efficient. Both Figs.~\ref{fig:2DPlotAlphavs2JM} and \ref{fig:2DPlotAlphavs4JM} show the data and signal distributions for events with $\mjjjj > 1.6\TeV$. 

\section{Signal modeling}
\label{sec:SignalSim}

\begin{figure}[htbp!]
\centering
\includegraphics[width=0.495\textwidth]{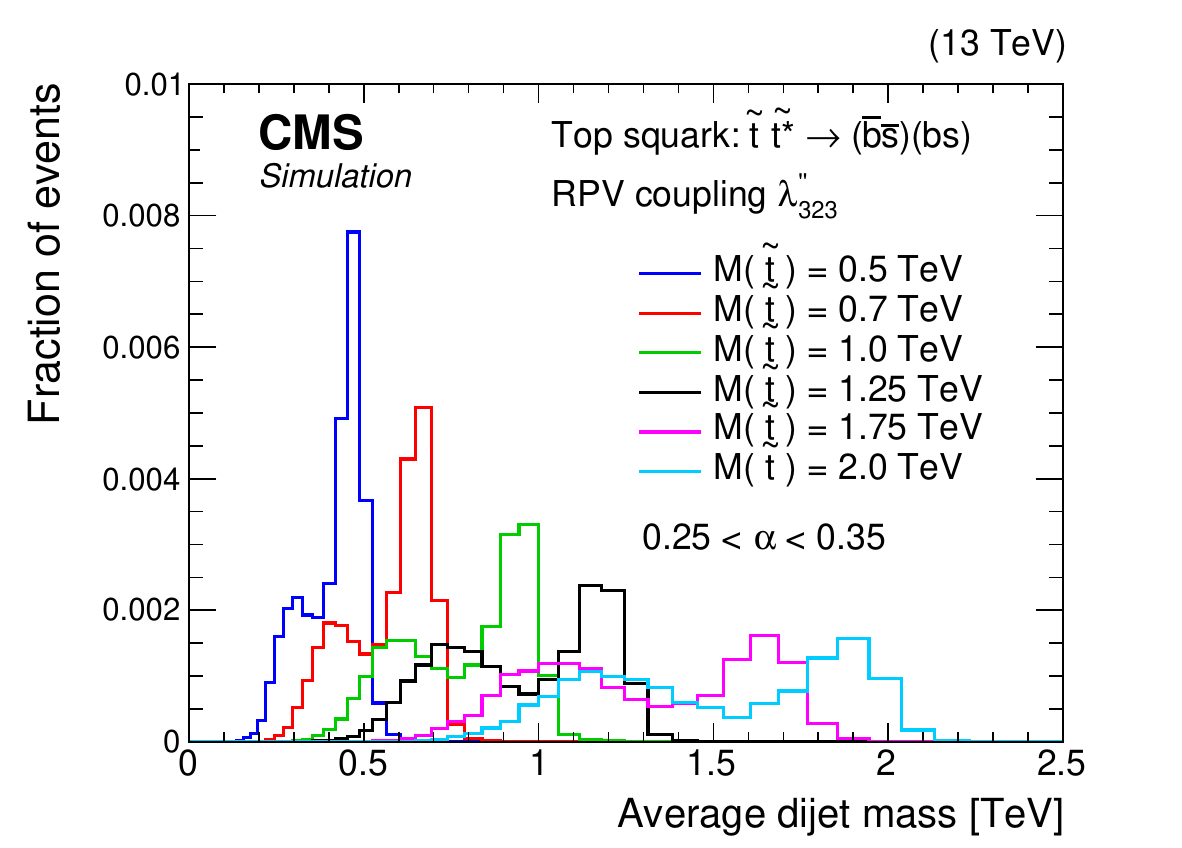}
\includegraphics[width=0.495\textwidth]{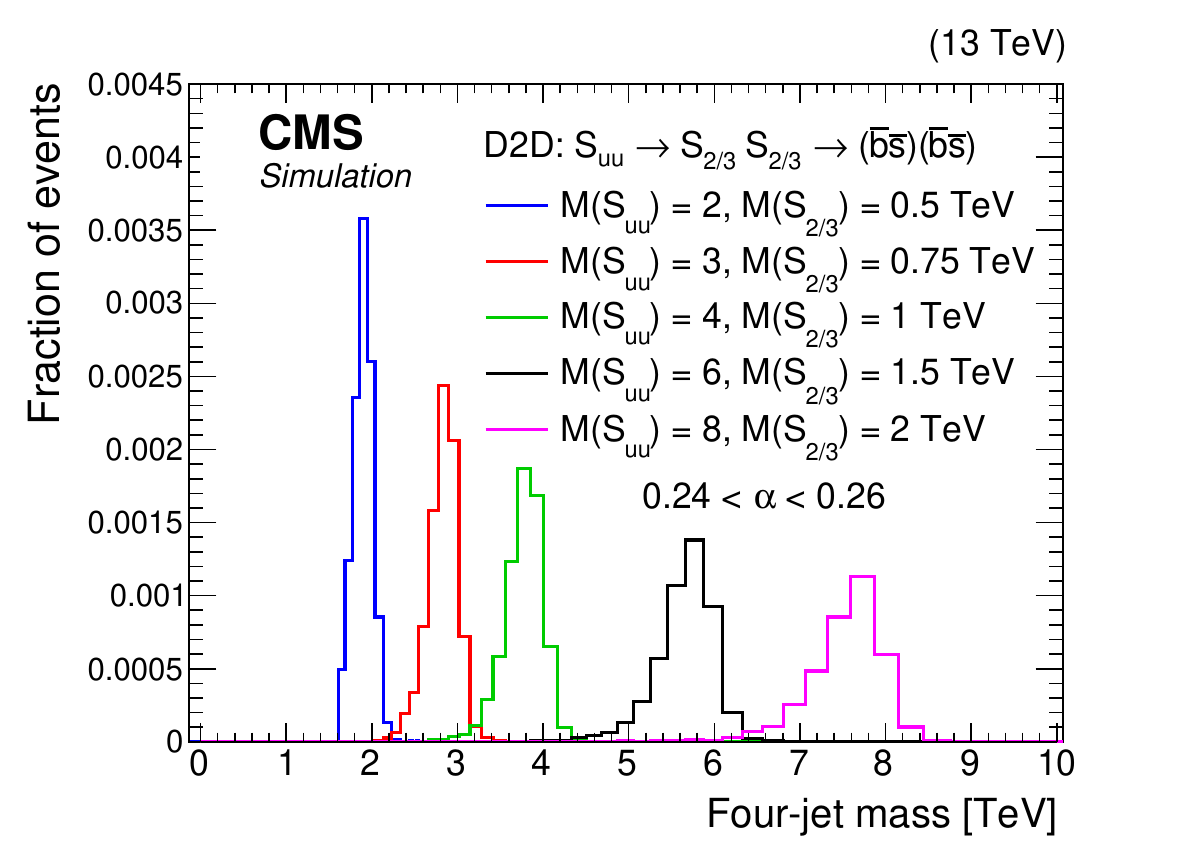} 
\caption{(Left) The average dijet mass distributions for different $R$-parity-violating top squark signal masses, with the kinematic and \PQb tagging selections applied. (Right) The four-jet mass distributions for different signal \PSuu diquark masses with $\atrue = 0.25$, with the kinematic and \PQb tagging selections applied. Shapes are shown for the $\alpha$ bins with the largest yield for each search, and have been normalized to have a unit area.
}
\label{fig:SignalShapes}
\end{figure}

Figure~\ref{fig:SignalShapes} shows the \mjj distributions for various RPV top squark signal masses (left), and the \mjjjj distributions for various \PSuu signal masses with $\atrue = M(\PStwothree)/M(\PSuu) = 0.25$ (right), in the $\alpha$ bins with the largest yield for both searches. The nominal kinematic and \PQb tagging selections are applied, as described in Section~\ref{sec:EventReconstruction}. For the resonant search, the peak of the \mjjjj distribution lies just below the \PSuu mass and also has a long tail to low \mjjjj resulting from final-state radiation, especially evident at higher resonance masses. The nonresonant signal shapes contain two peaks, one originating from the correct pairing of the four jets, and a second peak originating from an incorrect pairing. The incorrect dijet pairing peak becomes more pronounced at higher top squark masses. As the top squark mass increases, the resulting decay jets tend to be more isotropically distributed, reducing the effectiveness of the pairing algorithm in identifying the correct jet pairs by minimizing the $\Delta \mathrm{R}$.

The pairing efficiency is estimated by evaluating the fraction of events for which the reconstructed average dijet mass agrees with the true dijet resonance mass within three times the detector resolution of the reconstructed dijet mass. For the top squarks in the nonresonant search with $0.25 < \alpha < 0.35$, the pairing efficiency ranges from 1.0 for the lowest mass to 0.4 for the highest mass. For the D2D model in the  resonant search, the pairing efficiency is significantly higher, ranging from 1.0 at low diquark masses to 0.9 at the highest diquark masses with all alpha bins combined and for $\atrue = 0.25$. The four-jet resonance in the D2D model kinematically constrains the possible values of average dijet masses, resulting in a better pairing efficiency.

\begin{figure}[htbp!]
\centering
\includegraphics[width=0.495\textwidth]{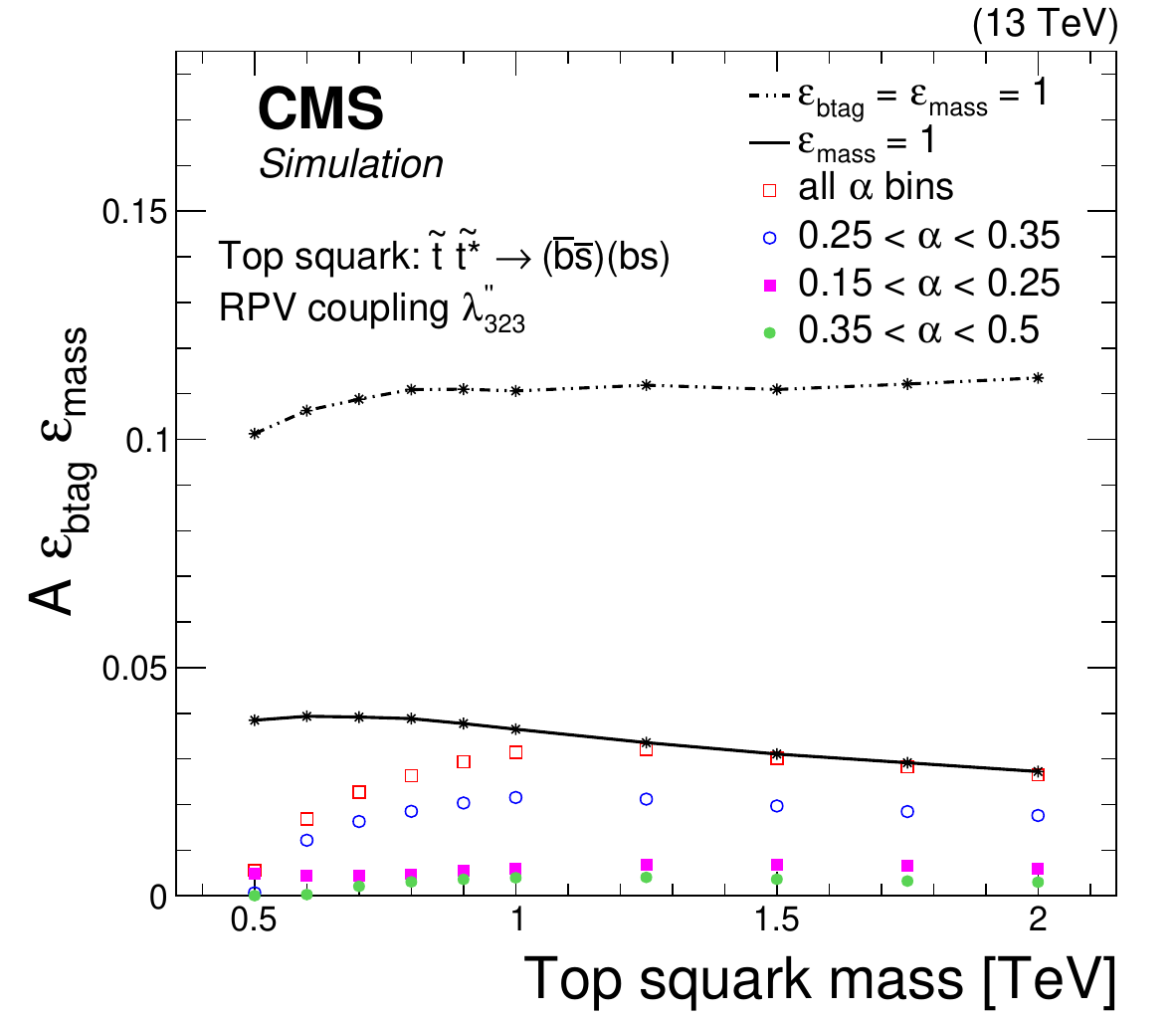}
\includegraphics[width=0.495\textwidth]{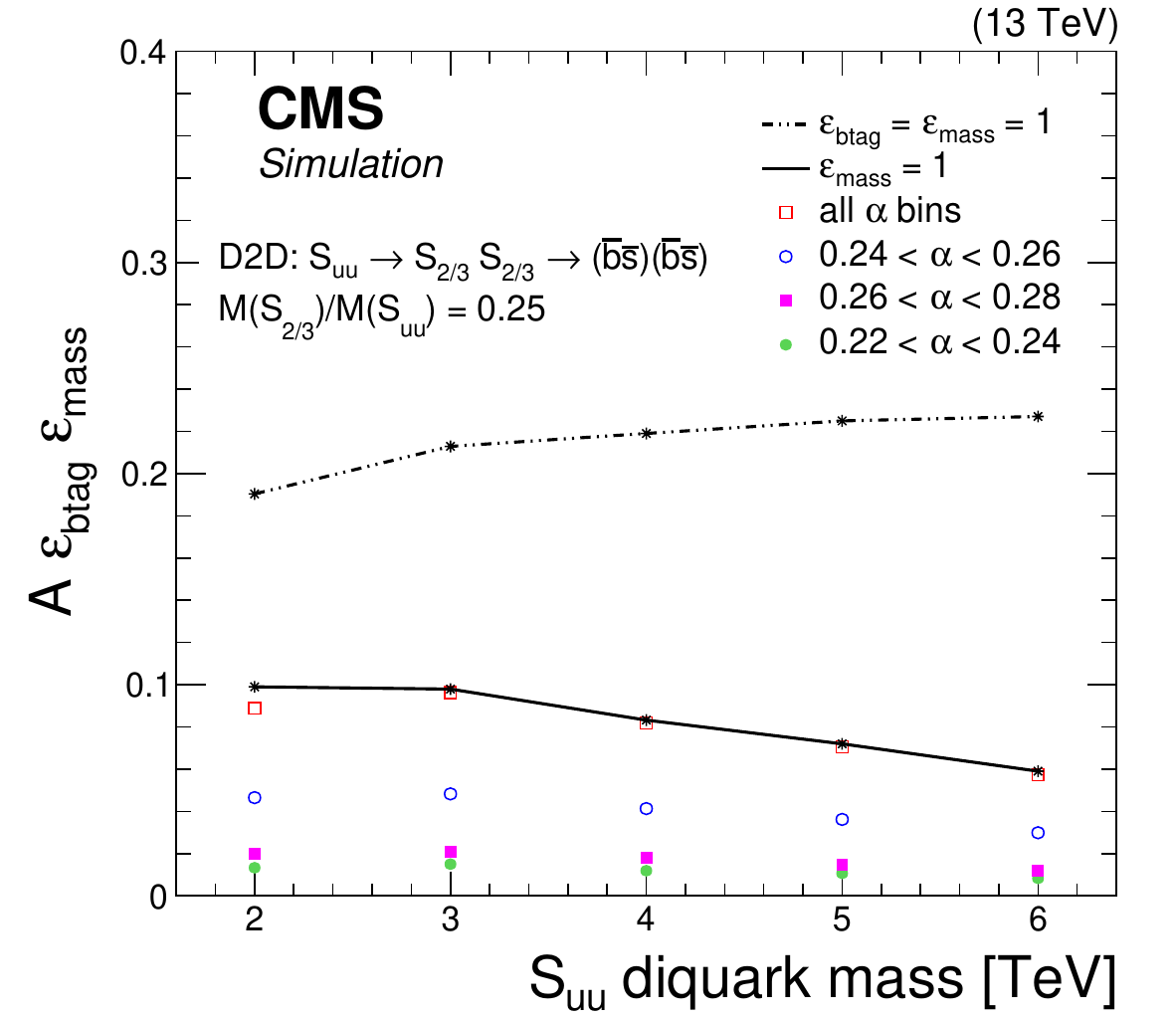}
\caption{The products of acceptance and efficiency for a nonresonant signal vs. the top squark mass (left), and a resonant signal vs. the diquark mass (right) inclusively, \ie, for all $\alpha$ values, and for the three $\alpha$ bins that contain the majority ($>$85\%) of the signal. The case where the efficiency of the mass selection is unity is shown as a solid line, and the case where both the mass selection and \PQb tagging efficiencies are unity is shown as a dashed black line.
}
\label{fig:AccXEffvsMass}
\end{figure}

Figure~\ref{fig:AccXEffvsMass} shows the product of acceptance and efficiency as a function of the top squark mass (left), and the \PSuu mass with $\atrue = 0.25$ (right). The acceptance ($A$) is defined as the fraction of events that pass the nominal kinematic selection, which corresponds to all requirements described in Section~\ref{sec:EventReconstruction},  except the \PQb tagging and the mass threshold requirements. The \PQb tagging efficiency ($\epsilon_{\PQb\text{tag}}$) is defined as the fraction of events that pass the \PQb tagging requirements with no $\alpha$ requirements applied. The mass efficiency ($\epsilon_{\text{mass}}$) is defined as the fraction of events that pass the \mjj (nonresonant) and \mjjjj (resonant) mass threshold requirements, as described in Section~\ref{sec:AnalysisStrategy}. Figure~\ref{fig:AccXEffvsMass} also shows the signal acceptance, when no mass threshold, no \PQb tagging requirements, and no $\alpha$ requirements are imposed, and only the nominal kinematic requirements are applied ($\epsilon_{\PQb\text{tag}} = \epsilon_{\text{mass}}$ = 1), no mass threshold and no $\alpha$ requirements are imposed ($\epsilon_{\text{mass}}$ = 1), and the signal times efficiency when no $\alpha$ requirements are imposed (all $\alpha$ bins).

For the nonresonant signal, Fig.~\ref{fig:AccXEffvsMass} (left) shows the signal acceptance times efficiency for three $\alpha$ slices, where the $0.25 < \alpha < 0.35$ slice has the highest acceptance. For the resonant search, Fig.~\ref{fig:AccXEffvsMass} (right) shows the signal acceptance times efficiency for the three $\alpha$ slices where approximately 80--85\% of the signal resides. As is expected for a signal with $\atrue = 0.25$, the $0.24 < \alpha < 0.26$ slice has the highest acceptance. For higher masses, a decreasing \PQb tagging efficiency leads to decreasing signal acceptance, for both resonant and nonresonant signals. This can be seen by comparing the line corresponding to $\epsilon_{\PQb\text{tag}} = \epsilon_{\text{mass}}$ = 1 (dashed black line) with the line corresponding to $\epsilon_{\text{mass}}$ = 1 (solid black line), in Fig.~\ref{fig:AccXEffvsMass}. Additionally, the acceptance is further reduced due to the mass asymmetry requirements, rejecting events with incorrectly paired jets, which occur more frequently at higher masses.

\section{Background estimation}
\label{sec:BackgroundEstimation}

{\tolerance=800
In this search, the background is estimated entirely on data. As in previous paired dijet searches \cite{AlePaper, DijetPaper, dijet13TeV}, we fit empirical functional forms to the observed \mjj distribution in the nonresonant search, and the observed \mjjjj distribution in the resonant search, using a background-only, extended, binned maximum likelihood. We perform these fits independently for all bins in $\alpha$, and use the following parametric functions: the dijet function (Dijet-3p) given by ${\rd\sigma}/{\rd m} =
{p_0(1-x)^{p_1}}/{x^{p_2}}$,  the modified dijet function (ModDijet-3p) given by ${\rd\sigma}/{\rd m} =
{p_0(1-x^{1/3})^{p_1}}/{x^{p_2}}$, and the power-law times
exponential function (PowExp-3p) given by ${\rd\sigma}/{\rd m} =
{p_0 \re^{-p_1(x)}}/{x^{p_2}}$,
where $p_0$, $p_1$, and $p_2$ are freely floating parameters, $x=m/\sqrt{s}$, and $m$ is the mass observable used for the search; in the resonant search this is \mjjjj and in the nonresonant search it is \mjj. All three functional forms are taken into account through the discrete profiling method~\cite{Dauncey:2015xga}, and the choice of background parameterization is treated as a systematic uncertainty.
\par}

\begin{figure}[ht]
    \centering
    \includegraphics[width=0.9\textwidth]{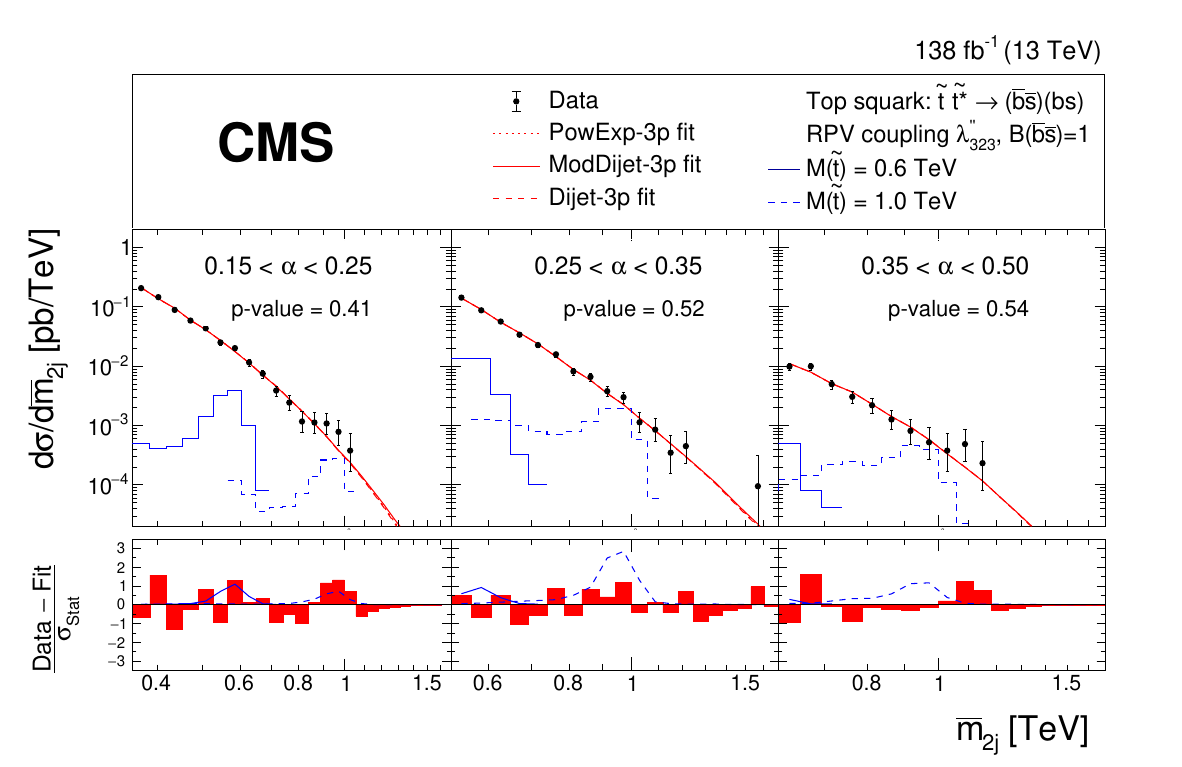}
    \caption{(Upper panel) The measured \mjj distributions (points) and background fits (red curves) in each slice of $\alpha$, in the nonresonant search. We note that the three background fits lie nearly on top of one another. (Lower panel) The residual difference between data and the Dijet-3p function, divided by the statistical
uncertainty of data ($\sigma_{\text{Stat}}$). In both panels, examples of predicted top squark pair production signals are shown, with cross sections equal to the observed upper limits at 95\% confidence level, for top squark masses of 0.6\TeV (blue solid line) and 1.0\TeV (blue dashed line).
We quote the $p$-value for each fit of the Dijet-3p function to the average dijet mass distribution in data, evaluated from pseudo-experiments.}
    \label{fig:RPV_BKG}
\end{figure}

\begin{figure}[ht]
    \centering
    \includegraphics[width=0.9\textwidth]{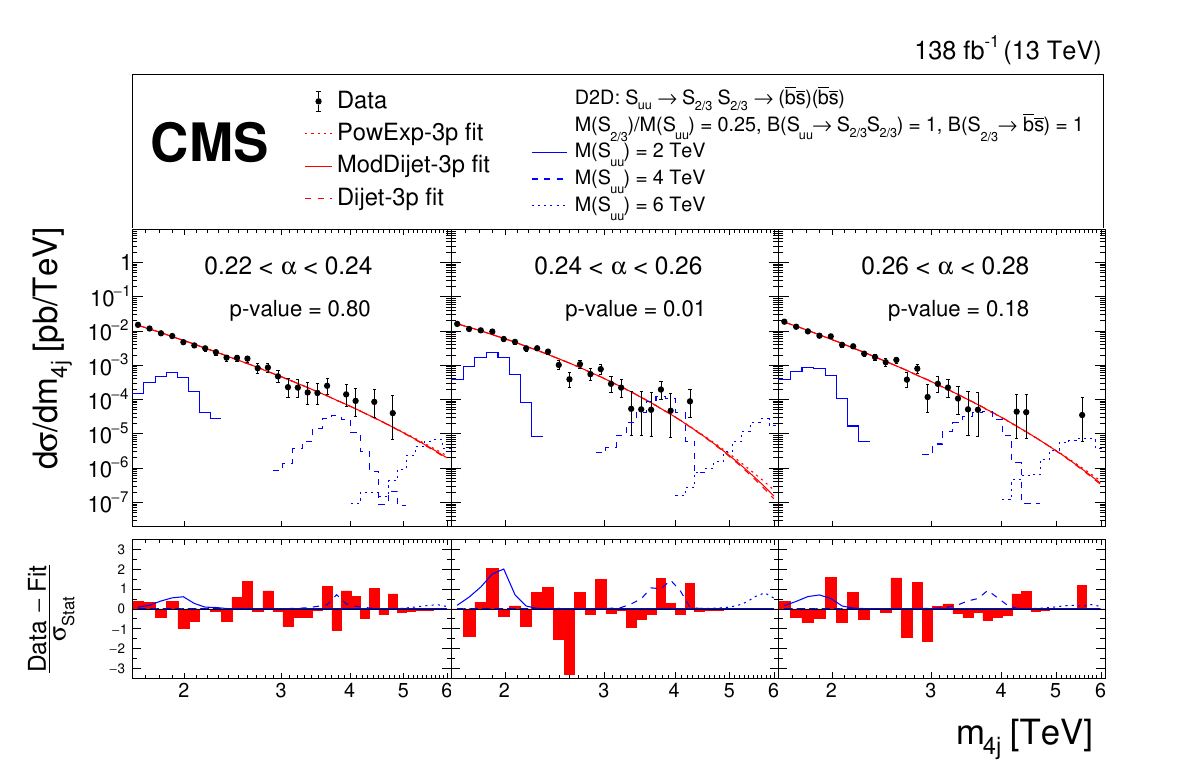}
    \caption{(Upper panel) The measured \mjjjj distributions (points) and background fits (red curves) in three out of the twelve slices of $\alpha$ in the resonant search. We note that the three background fits lie nearly on top of one another. (Lower panel) The residual difference between data and the Dijet-3p function, divided by the statistical
uncertainty of data ($\sigma_{\text{Stat}}$). In both panels, examples of predicted D2D signals are shown, with cross sections equal to the observed upper limits at 95\% confidence level, for \PSuu diquark masses of 2\TeV (blue solid line), 4\TeV (blue dashed line), and 6\TeV (blue dotted line), and for $M(\PStwothree)/M(\PSuu) = 0.25$. We quote the $p$-value for each fit of the Dijet-3p function to the four-jet mass distribution in data, evaluated from pseudo-experiments.}
    \label{fig:SUU_BKG}
\end{figure}

\begin{figure}[ht]
    \centering
    \includegraphics[width=0.8\textwidth]{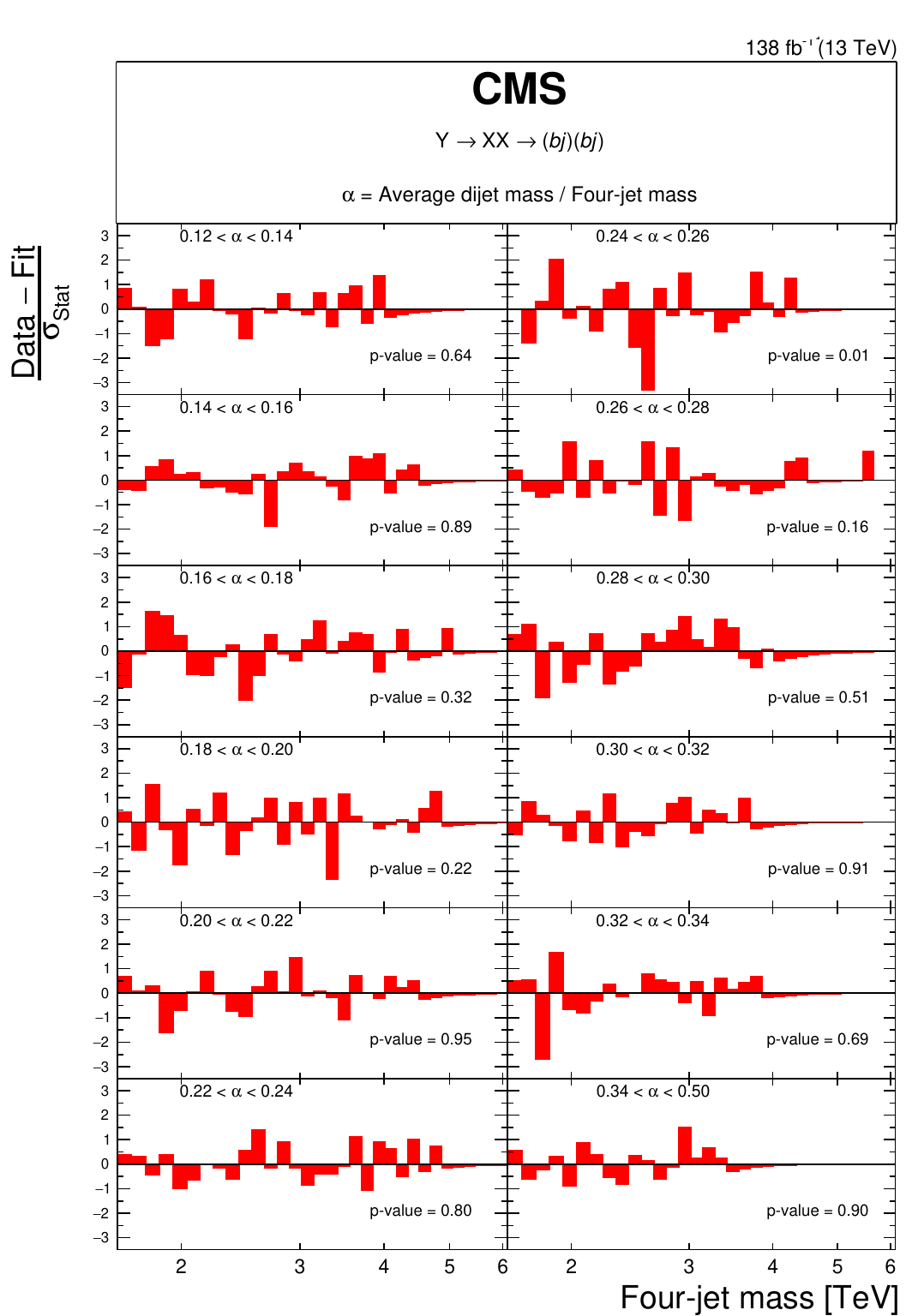}
    \caption{The residual difference between data and the Dijet-3p function, divided by the statistical uncertainty of data ($\sigma_{\text{stat}}$), in all twelve $\alpha$ slices used in the resonant search. We quote the $p$-value for each fit of the Dijet-3p function to the four-jet mass distribution in data, evaluated from pseudo-experiments. The distributions and fits for the $0.22 < \alpha < 0.24$, $0.24 < \alpha < 0.26$, and $0.26 < \alpha < 0.28$ bins are shown in Fig.~\ref{fig:SUU_BKG}.}
    \label{fig:ALL_SUU_BKG}
\end{figure}

Figure~\ref{fig:RPV_BKG} shows the three functional forms fitted to the observed differential cross section, as functions of \mjj, for each of the three $\alpha$ slices used in the nonresonant search.  Similarly, Fig.~\ref{fig:SUU_BKG} shows the same functional forms fitted to the observed differential cross section as functions of \mjjjj, in three of the twelve $\alpha$ slices used in the resonant search. Figure~\ref{fig:ALL_SUU_BKG} shows the residual difference between the data and the fit of the \mjjjj distribution with the Dijet-3p function, divided by the statistical uncertainty, in all twelve $\alpha$ slices used in the resonant search. The $p$-values~\cite{GrossVitells} quoted in Fig.~\ref{fig:RPV_BKG},~\ref{fig:SUU_BKG} and~\ref{fig:ALL_SUU_BKG} are evaluated conducting a goodness-of-fit test on pseudo-experiments, utilizing a binned likelihood ratio with respect to the saturated
model as the test statistic. These $p$-values and the Fisher tests~\cite{Ftest}, all indicate that the background functions provide a good description of the data.

\clearpage

\section{Systematic uncertainties}
\label{sec:Systematics}

For both the resonant and nonresonant searches, systematic uncertainties are implemented as nuisance parameters in the likelihood model, with Gaussian constraints for the jet energy scale and resolution, and log-normal constraints for the integrated luminosity and \PQb tagging uncertainty. A multibin counting experiment likelihood is used for signal extraction, which is a product of Poisson distributions corresponding to different bins. The likelihood is independently evaluated at each resonance mass value. 

The primary source of the uncertainty in the background modeling comes from the choice of the functional form used to describe the background and the range of its associated parameter values. The parameters of the empirical function are treated as unconstrained nuisance parameters and are determined through profiling. To incorporate the uncertainty related to the choice of the functional form itself, the discrete profiling method is applied. This technique treats the functional form as a discrete nuisance parameter and profiles it in a manner analogous to continuous nuisance parameters. These uncertainties have an effect on the extracted cross section limits that range from a few percent up to 50\%, depending on the mass. 

The systematic uncertainties propagated into the signal prediction are due to the jet energy scale, jet energy resolution, integrated luminosity, and \PQb tagging. The systematic uncertainty in the signal is dominated by the \PQb tagging uncertainty, which ranges from 6 to 33\%, depending on the signal mass, and is applied as a single uncertainty value in the final event yield. The jet energy scale uncertainty is below 2\% across all \mjj and \mjjjj values. This uncertainty is accounted for by shifting the signal mass distributions  by $\pm$2\%. Similarly, the jet energy resolution uncertainty leads to a 10\% uncertainty in the resolution of the resonance masses and is included in the statistical model by examining the impact of increasing the reconstructed signal mass widths by 10\%. Additionally, the total integrated luminosity has an uncertainty of 1.6\%~\cite{CMS-LUM-17-003,CMS-PAS-LUM-17-004,CMS-PAS-LUM-18-002}. At lower masses, the uncertainties related to the background modeling have the largest impact on the extracted cross section limit and signal significance, while the uncertainties related to the signal modeling have a relatively smaller impact. On the other hand, at higher masses, where the event count in data becomes limited, the statistical uncertainty of the data and the uncertainties in the signal modeling have the greatest impact.

\section{Results}
\label{sec:Results}

\begin{figure}[htbp!]
    \centering
    \includegraphics[width=0.65\textwidth]{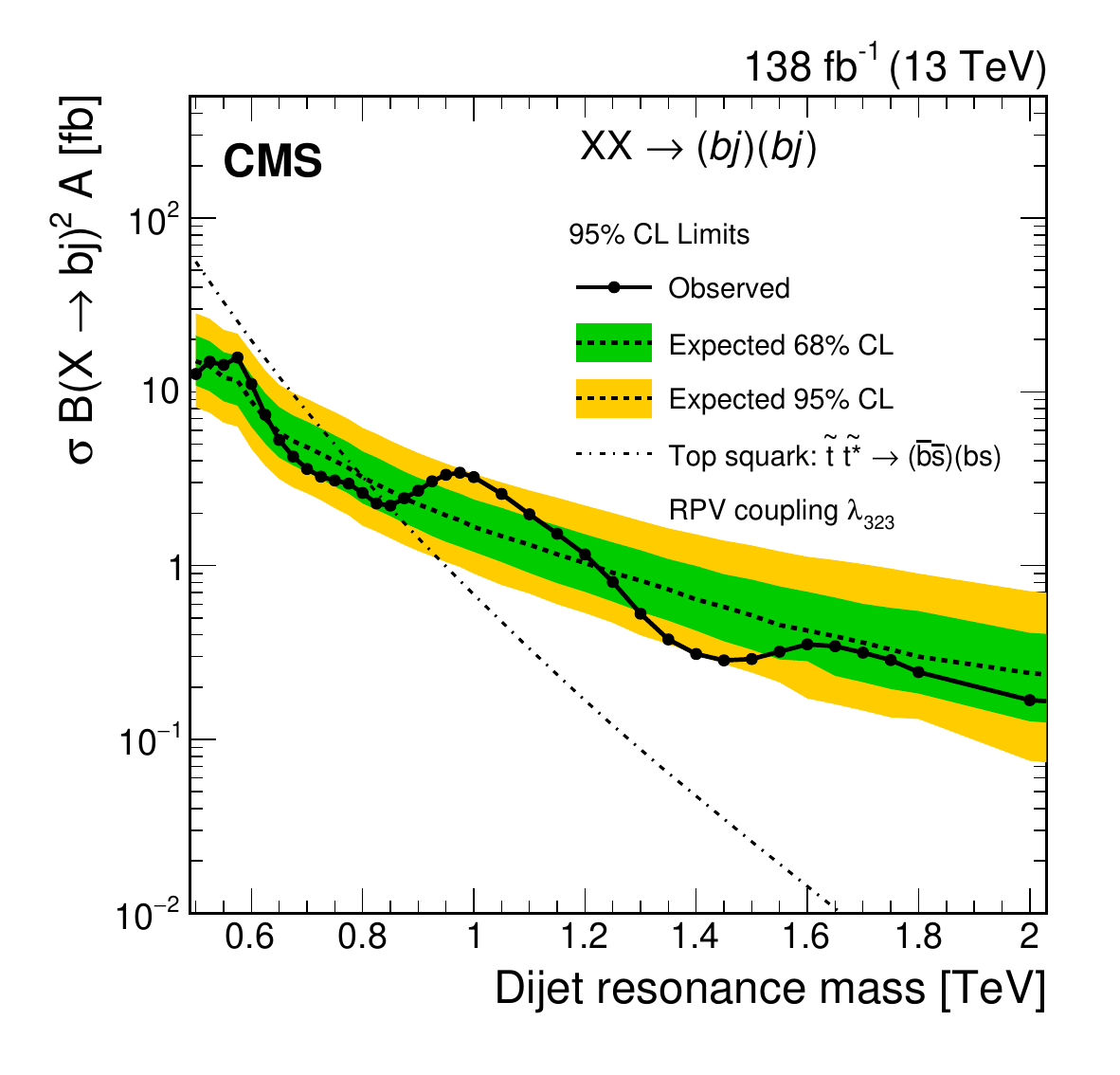}
    \caption{The observed 95\% \CL upper limits (black line with points) on the product of the cross section, branching fraction, and acceptance in the nonresonant search for $\PX \PX \to (\PQb \Pj)(\PQb \Pj)$.  The expected limits (dashed line) and their variations at the 68\% and 95\% \CL (shaded bands) are also shown. Limits are compared to the predicted cross section of the RPV model~\cite{Borschensky:2014cia,Beenakker:2016lwe} (dot-dashed line), assuming $\mathcal{B}(\PSQt \to \PAQb\PAQs) = 1$.}
    \label{fig:RPVlimits}
\end{figure}

\begin{figure}[htbp!]
    \centering
    \includegraphics[width=0.65\textwidth]{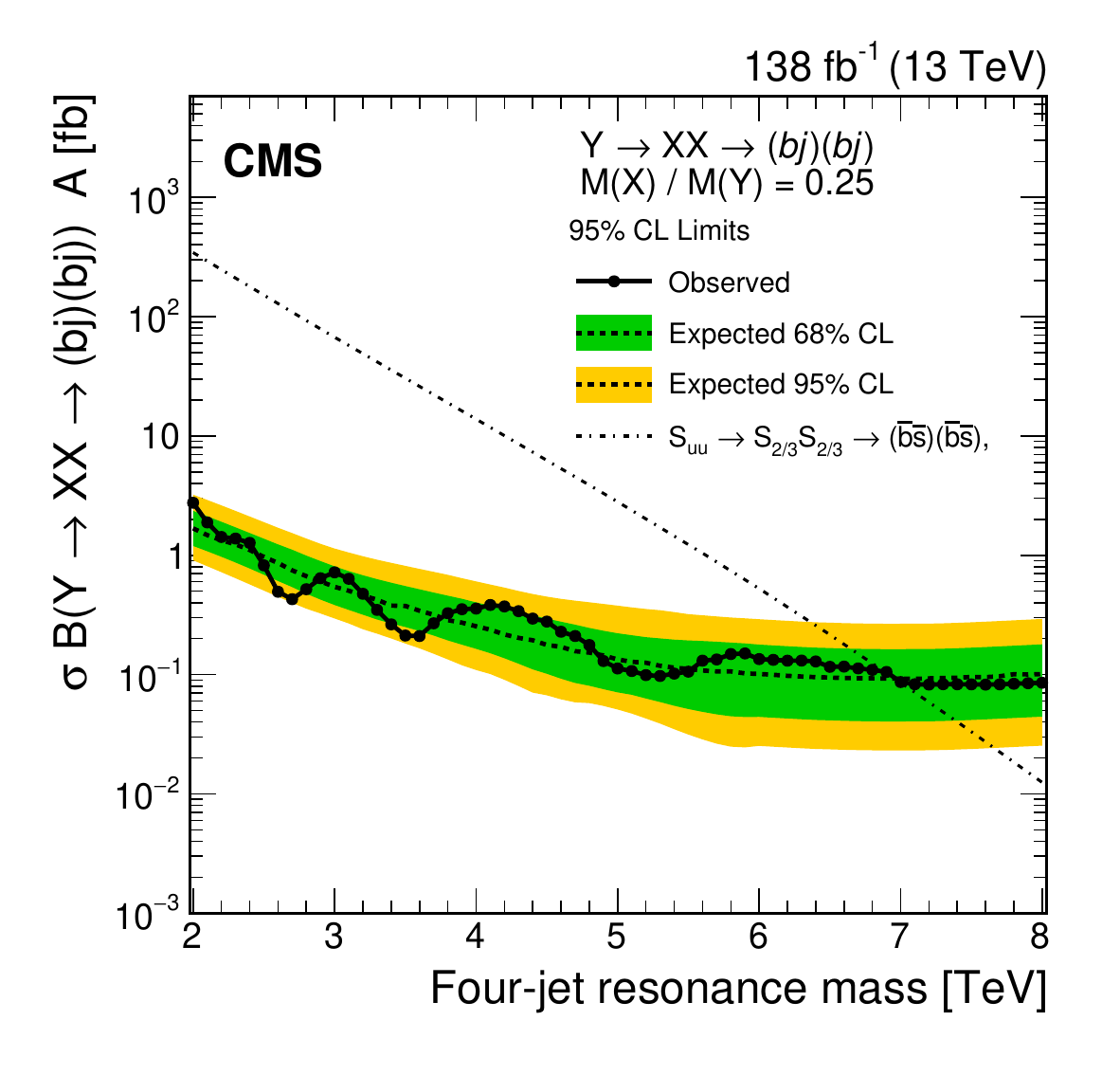}
    \caption{The observed 95\% \CL upper limits (black line with points) on the product of the cross section, branching fraction, and acceptance in the resonant search for $\PY \to \PX \PX \to (\PQb \Pj)(\PQb \Pj)$ with $\atrue = M(\PX)/M(\PY) = 0.25$. The expected limits (dashed line) and their variations at the 68\% and 95\% \CL (shaded bands) are also shown. Limits are compared to the predicted cross section of the D2D diquark model~\cite{Bogdanpaper} (dot-dashed line) for $M(\PStwothree)/M(\PSuu) = 0.25$ and $\mathcal{B}(\PSuu \to \PStwothree\PStwothree \to (\PAQb\PAQs)(\PAQb\PAQs)) = 1$.}
    \label{fig:Suulimits}
\end{figure}

\begin{figure}[htbp!]
    \centering
    \includegraphics[width=0.85\textwidth]{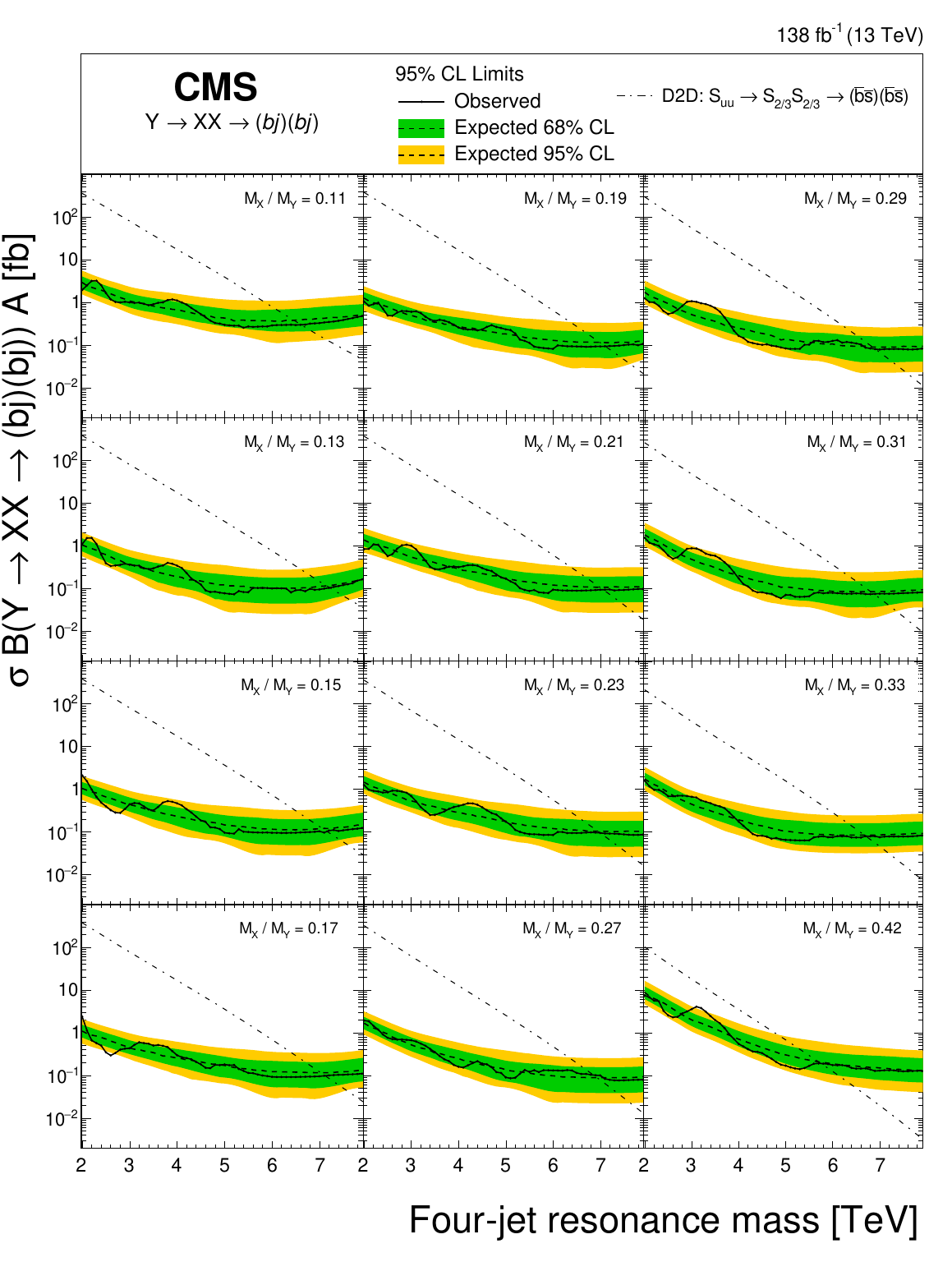}
    \caption{The observed 95\% \CL upper limits (black line with points) on the product of the cross section, branching fraction, and acceptance in the resonant search for $\PY \to \PX \PX \to (\PQb \Pj)(\PQb \Pj)$ for the other twelve values of $\atrue = M(\PX)/M(\PY)$. The expected limits (dashed line) and their variations at the 68\% and 95\% \CL (shaded bands) are also shown. Limits are compared to the predicted cross section of the D2D diquark model~\cite{Bogdanpaper} (dot-dashed line) for the respective \atrue values, assuming $\mathcal{B}(\PSuu \to \PStwothree\PStwothree \to (\PAQb\PAQs)(\PAQb\PAQs)) = 1$.}
    \label{fig:AllLimits}
\end{figure}

We use the asymptotic approximation~\cite{JUNK1999435,ALRead2002,Cowan:2010js,ATLAS:2011tau} of the test statistic to set 95\% confidence level (\CL) upper limits on signal cross section. The following results have been determined using the CMS statistical analysis tool \textsc{Combine}~\cite{CMS:2024onh}. Figures~\ref{fig:RPVlimits} and~\ref{fig:Suulimits} show the 95\% \CL upper limits on the product of the cross section ($\sigma$), the branching fraction ($\mathcal{B}$), and the acceptance ($A$) 
as functions of the true dijet resonance mass for both nonresonant and resonant searches. In the latter case, the limits corresponding to $\atrue = M(\PX)/M(\PY) = 0.25$ are shown. The branching fraction ($\mathcal{B}$) is defined for the decay of the dijet resonance to a \PQb quark and a light quark, and is set to 1, as described in Section~\ref{sec:SimulatedDataSamples}. The acceptance ($A$) in Figs.~\ref{fig:RPVlimits} and~\ref{fig:Suulimits} is the same as defined in Section~\ref{sec:SignalSim}, and does not include the \PQb tagging and trigger efficiency, such that the results presented are model independent as well as independent of detector-specific effects. Additionally, Fig.~\ref{fig:AllLimits} shows the results for all other \atrue values in the resonant search,  including for $\atrue = 0.11$. Even though the resonant search is performed for $\alpha$ bins starting at 0.12, sensitivity to a signal with $\atrue = 0.11$ is still retained, since a sizable fraction of signal events is found in $\alpha > 0.12$. A separate statistical fit is performed for each signal mass hypothesis. For the nonresonant search, it is evaluated in steps of 0.025\TeV between 0.5 and 1.0\TeV, steps of 0.05\TeV between 1.0 and 1.8\TeV, and steps of 0.2\TeV between 1.8 and 2.0\TeV. For the resonant search, it is evaluated from 2.0 to 8.0\TeV in steps of 0.1\TeV. The step sizes are chosen such that they are commensurate with the experimental mass resolution. Signal is excluded at 95\% \CL in the mass regions where the theoretical prediction exceeds the observed upper limit for a given model. 

As shown in Fig.~\ref{fig:RPVlimits}, this analysis exclude at 95\% \CL top squarks with masses in the ranges 500--850\GeV (observed) and 500--800\GeV (expected), for top squarks decaying via $R$-parity-violating trilinear coupling with strength $\lambda''_{323}$. This extends the prior best mass limits~\cite{AlePaper} that excluded the mass regions 80--270, 285--340, and 400--525\GeV.

Based on Figs.~\ref{fig:Suulimits} and~\ref{fig:AllLimits} we exclude \PSuu diquarks at 95\% \CL with masses in the ranges 2.0--6.6\TeV (observed) and 2.0--6.4\TeV (expected), for an \atrue value of 0.11. For an \atrue value of 0.42, we exclude masses in the ranges 2.0--5.8\TeV (observed) and 2.0--5.6\TeV (expected). The 95\% \CL lower bound on the \PSuu diquark masses, for \atrue values between 0.13 and 0.33, varies between 7.2 and 6.7\TeV (observed) and between 7.1 and 6.6\TeV (expected). These are the first limits on the D2D model~\cite{Bogdanpaper}. 

\begin{figure}[ht]
    \centering
    \includegraphics[width=0.65\textwidth]{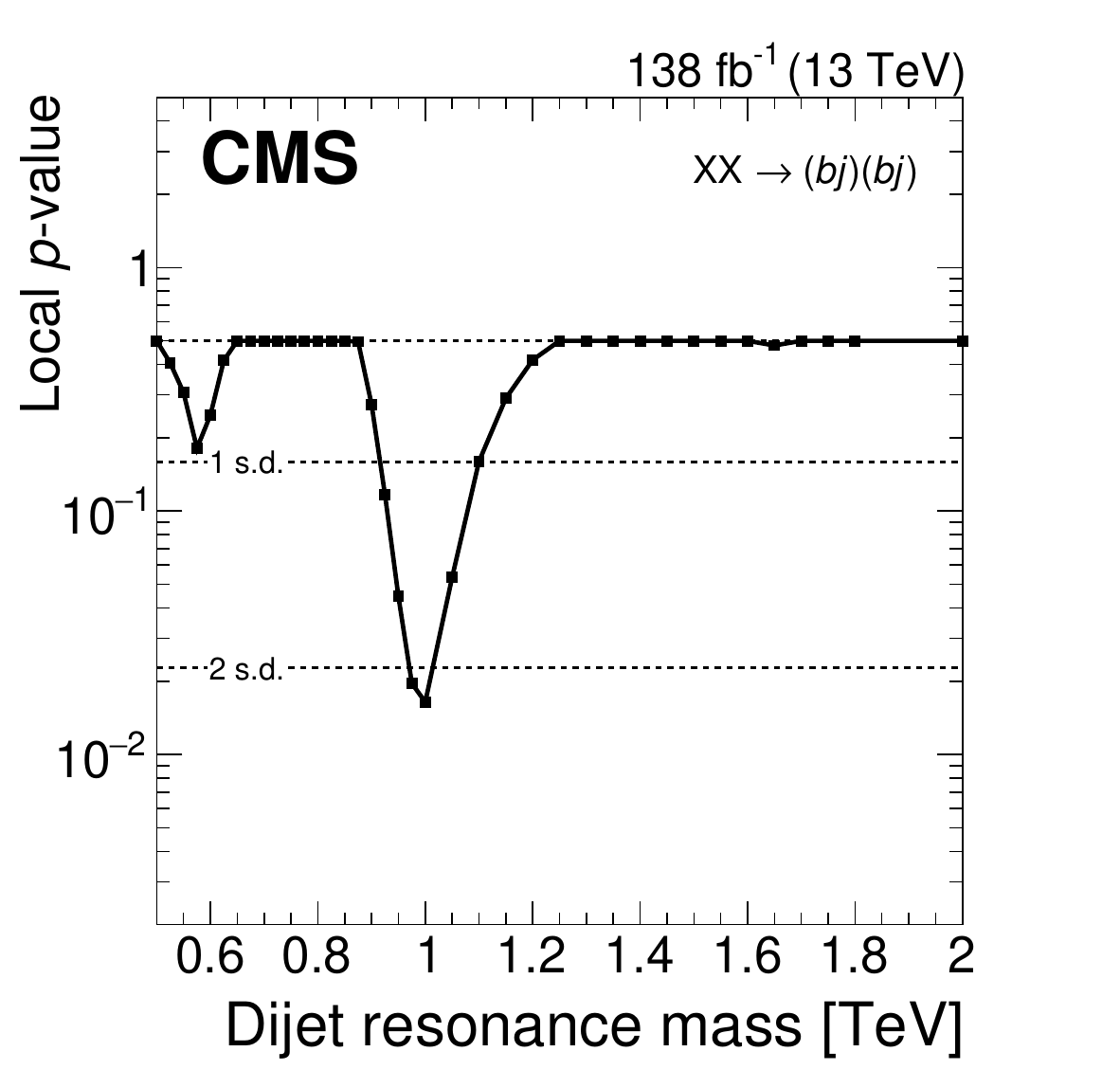}
    \caption{The observed local $p$-value for a signal across all  $\alpha$ bins for the nonresonant search $\PX \PX \to (\PQb \Pj)(\PQb \Pj)$. Dashed lines indicate the corresponding levels of local significance, expressed in standard deviations (s.d.).}
    \label{fig:RPV_pvalues}
\end{figure}

\begin{figure}[htbp!]
    \centering
    \includegraphics[width=0.85\textwidth]{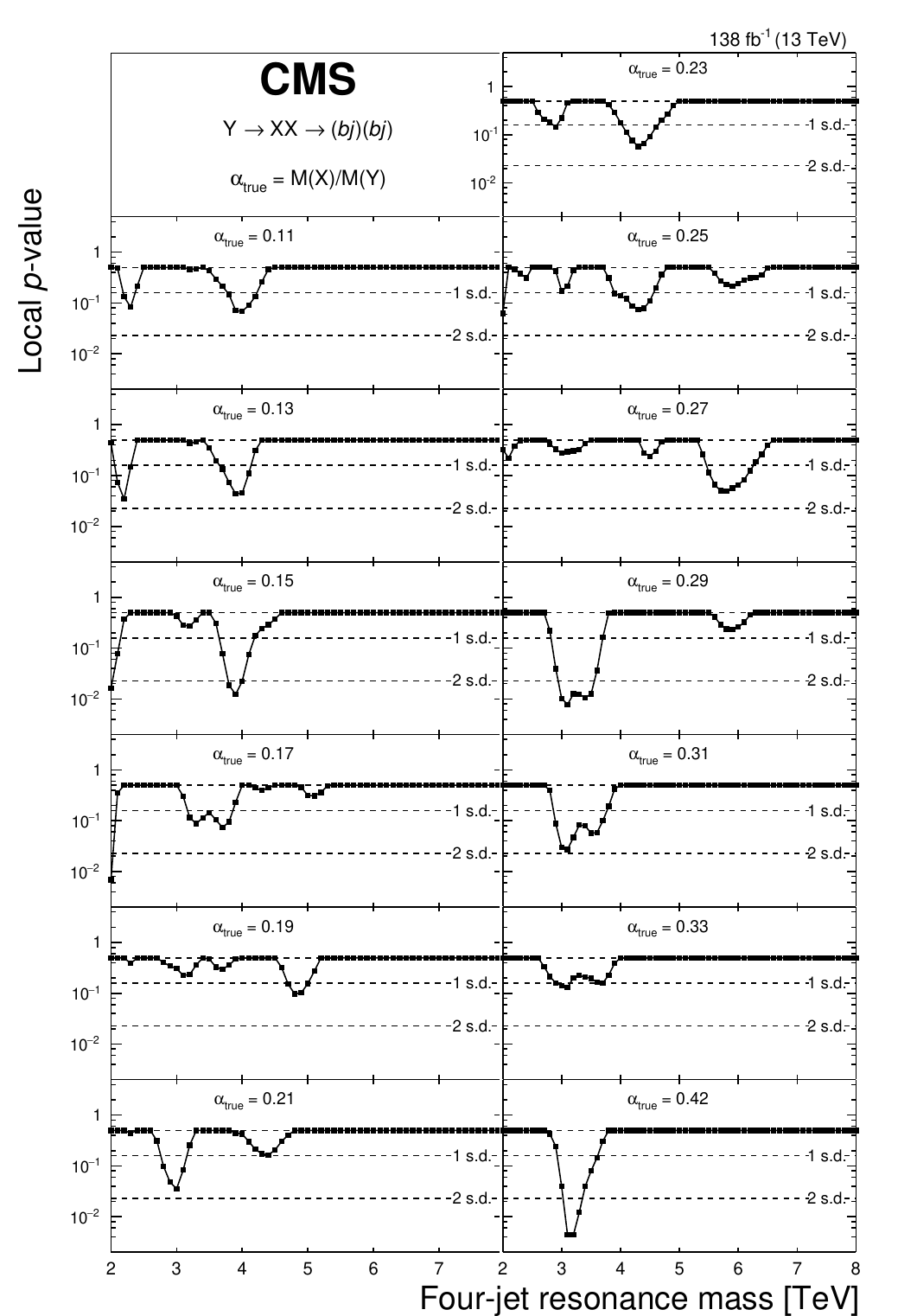}
    \caption{The observed local $p$-values for a signal across all $\alpha$ bins for the resonant search $\PY \to \PX \PX \to (\PQb \Pj)(\PQb \Pj)$ for 13 values of $\atrue = M(\PX)/M(\PY)$. Dashed lines denote the corresponding levels of local significance, expressed in standard deviations (s.d.).}
    \label{fig:Suu_all_pvalues}
\end{figure}

Figure~\ref{fig:RPV_pvalues} shows the corresponding $p$-values for dijet resonance signals for the nonresonant search, as functions of the resonance mass, with the corresponding local significance expressed in s.d., included for reference. We observe a local significance of 2.1 s.d.~at a dijet resonance mass of 1.0\TeV, corresponding to an observed cross section of $1.68 \pm 0.90\unit{fb}$. We note that the previous search that did not use \PQb tagging~\cite{DijetPaper} observed an excess with 3.6 s.d.~local significance at a compatible dijet resonance mass of 0.95\TeV, which corresponded to a best fit cross section of $7.95 \pm 2.52\unit{fb}$. Figure~\ref{fig:Suu_all_pvalues} shows the observed $p$-values for all \atrue values in the resonant search. The most significant signal hypothesis for the resonant search with a 2.6 s.d.~effect occurs for $\atrue = M(\PX)/M(\PY) = 0.42$ for a four-jet resonance mass of 3.1\TeV, as seen in Fig.~\ref{fig:Suu_all_pvalues}. In addition, Fig.~\ref{fig:Suu_all_pvalues} also shows that for $\atrue = 0.29$ and four-jet resonance masses of $3.0 < M(\PY) < 3.5\TeV$, there is a 2.3 s.d.~excess. We note this in order to compare it with effects observed in an extension of the previous search that did not use flavor tagging~\cite{DijetPaper}. In the previous untagged search, the local significance for $\atrue = 0.29$ in the compatible mass region $3.5 < M(\PY) < 3.6\TeV$ was 3 s.d.~for a comparably narrow resonance of fractional natural width $\Gamma/M=1.5\%$ and increased to 3.9 s.d.~for a broad resonance with $\Gamma/M = 10\%$. Despite the fact that the nonresonant and resonant analyses are looking for different signals, the observed limits are highly correlated since they are probing the same events for a localized excess. Approximately 36\% of the events in the resonant search region of $0.28 < \alpha < 0.30$ and $2.5 < \mjjjj < 3.5\TeV$, which accounts for the majority of the observed excess at $\atrue = 0.29$, are also selected by the nonresonant search in the region $0.85 < \mjj < 1\TeV$ for $0.25 < \alpha < 0.35$.

\clearpage

\begin{figure}[htbp]
    \centering
    \includegraphics[width=\textwidth]{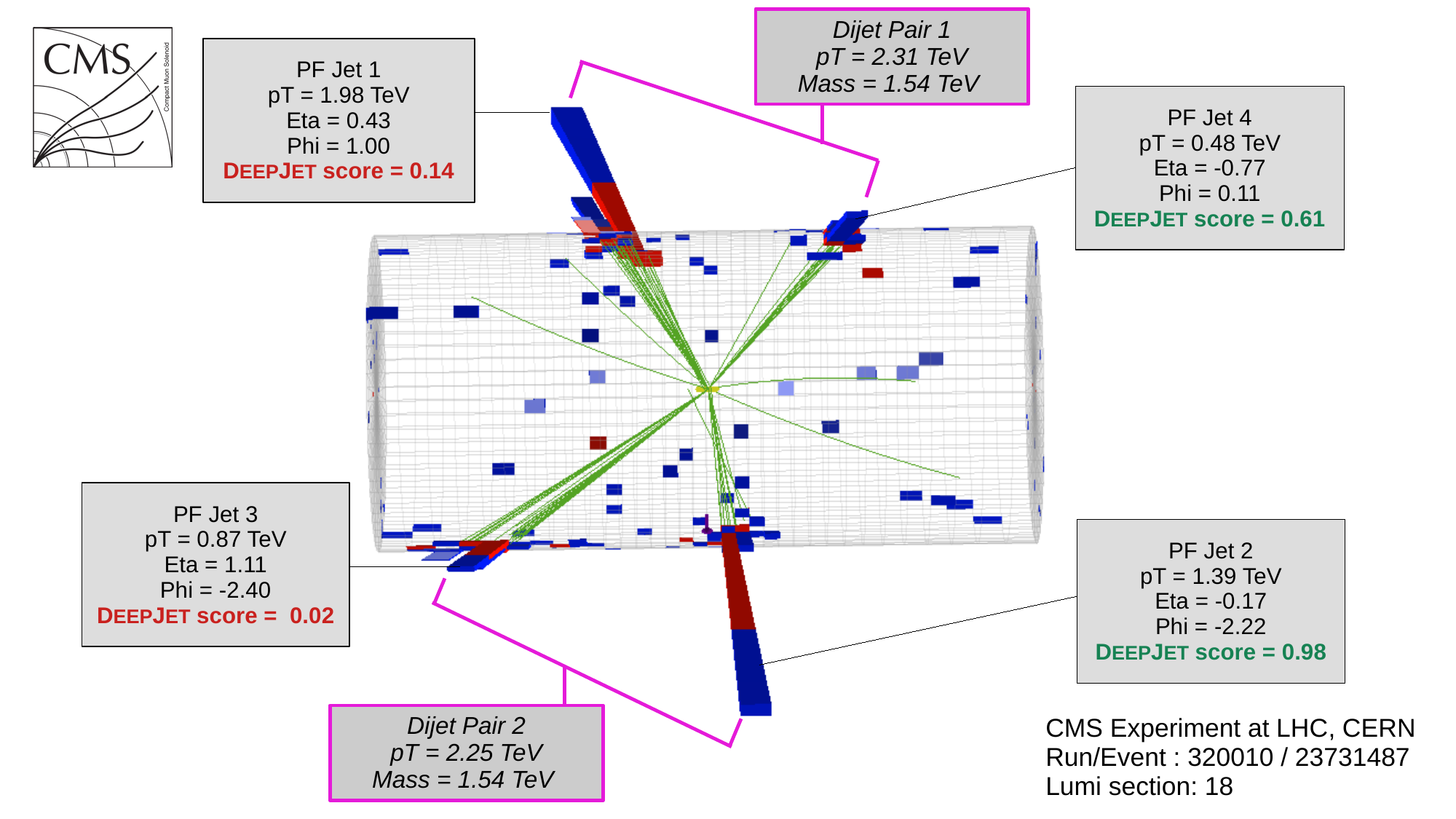}
    \caption{Three-dimensional display of the event with the highest $\mjjjj = 5.5\TeV$ and highest $\mjj =1.5\TeV$. The display shows the energy deposited in the electromagnetic (red) and hadron (blue) calorimeters, and the reconstructed tracks of charged particles (green). The two \PQb jets in the event have passing \PQb tagging scores, shown in green, and the two other jets in the event have failing \PQb tagging scores, shown in red, using the 2018 `medium' working point threshold of 0.2770.
    The grouping of the four observed jets into two dijet pairs (purple box) is discussed in the text.}
    \label{fig:EventDisplay}
\end{figure}

Figure~\ref{fig:EventDisplay} shows a display of the highest mass event observed, which was collected during 2018, with $\mjjjj = 5.5\TeV$ and $\mjj = 1.54\TeV$. The two dijet pairs are back-to-back in azimuthal angle ($\abs{\Delta \phi}$ = 3.1), nearby in pseudorapidity ($\abs{\Delta\eta}$ = 0.20), and have a mass asymmetry of 0.001. The $\Delta R_{1}$ and $\Delta R_{2}$ values between the two jets in the two dijet pairs are 1.5 and 1.3, respectively. Among the dijet pairs, jet 4 in dijet 1 and jet 2 in dijet 2 are \PQb tagged, and the \deepjet score, \pt, $\eta$, and $\phi$ of each jet can be seen in Fig.~\ref{fig:EventDisplay}. This event appears in the nonresonant search in Fig.~\ref{fig:RPV_BKG}, on the high-mass tail of the \mjj distribution for the $\alpha$ slice $0.25 < \alpha < 0.35$, and it appears in the resonant search in Fig.~\ref{fig:SUU_BKG}, on the high-mass tail of the \mjjjj distribution for the $\alpha$ slice $0.26 < \alpha < 0.28$. The effect of this event on the resonant search can be seen in Fig.~\ref{fig:Suu_all_pvalues}, for $\atrue = 0.27$ at a four-jet resonance mass of 5.8\TeV, where it produces a $p$-value of 0.05 corresponding to 1.6 s.d.

\section{Summary}
\label{sec:Summary}

A search for nonresonant and resonant production of pairs of dijet resonances, with a bottom (\PQb) jet within each dijet, has been performed. Data from proton-proton collisions at $\sqrt{s}=13\TeV$ are used in this search, collected by the CMS experiment at the LHC, corresponding to an integrated luminosity of \DatasetLumi. The analysis is conducted across steeply falling dijet (\mjj) and four-jet (\mjjjj) mass distributions, binned in $\alpha = \mjj/\mjjjj$. The mass distributions are fitted with a sum of empirical background functions and the simulated shapes of resonance signals, to search for a mass bump indicative of a heavy resonance. We observe no significant evidence for new resonances. 

Upper limits at 95\% confidence level (\CL) are provided for the production cross section multiplied by the branching fraction and acceptance, for models predicting pair-produced resonances decaying to pairs of quarks. These are the first LHC limits on resonant production of pairs of dijet resonances decaying to \PQb jets, in which all resonances considered are beyond the standard model. For nonresonant production, the limits are presented as functions of the mass of the spin-0 dijet resonance, in the range 0.5 to 2.0\TeV. For resonant production, the limits are given for values of four-jet resonance mass between 2 and 8\TeV.

Limits from the nonresonant search are compared to a model~\cite{Evans2012,Barbier:2004ez} of $R$-parity-violating supersymmetry, with pair-produced top squarks decaying to bottom and strange quarks. Top squarks with masses between 0.5 and 0.8\TeV are excluded at 95\% \CL. This significantly extends the mass limit on these top squark decays, which was previously about 0.5\TeV~\cite{AlePaper}.
The largest excess observed in the nonresonant search occurs for a dijet resonance mass of 1.0\TeV and has a local significance of 2.1 standard deviations.

Limits from the resonant search are compared 
to a recent model~\cite{Bogdanpaper} of massive intermediate scalar diquarks (\PSuu) decaying to pairs of final
state diquarks (\PStwothree), and exclude the \PSuu for masses between 2 and 7\TeV, for
nearly all values of the \PStwothree mass considered.  We note that these are the first limits on this model of \PSuu decays. The most significant excess seen in the resonant search occurs at an \PSuu mass of 3.1\TeV and an \PStwothree mass of 1.3\TeV, and has a local significance of 2.6 standard deviations.

\begin{acknowledgments}
We thank Bogdan Dobrescu for supplying the code and cross section for the D2D model and for his work developing a benchmark model specifically for this search.

We congratulate our colleagues in the CERN accelerator departments for the excellent performance of the LHC and thank the technical and administrative staffs at CERN and at other CMS institutes for their contributions to the success of the CMS effort. In addition, we gratefully acknowledge the computing centers and personnel of the Worldwide LHC Computing Grid and other centers for delivering so effectively the computing infrastructure essential to our analyses. Finally, we acknowledge the enduring support for the construction and operation of the LHC, the CMS detector, and the supporting computing infrastructure provided by the following funding agencies: SC (Armenia), BMFWF and FWF (Austria); FNRS and FWO (Belgium); CNPq, CAPES, FAPERJ, FAPERGS, and FAPESP (Brazil); MES and BNSF (Bulgaria); CERN; CAS, MoST, and NSFC (China); MINCIENCIAS (Colombia); MSES and CSF (Croatia); RIF (Cyprus); SENESCYT (Ecuador); ERC PRG and PSG, TARISTU24-TK10 and MoER TK202 (Estonia); Academy of Finland, MEC, and HIP (Finland); CEA and CNRS/IN2P3 (France); SRNSF (Georgia); BMFTR, DFG, and HGF (Germany); GSRI (Greece); MATE and NKFIH (Hungary); DAE and DST (India); IPM (Iran); SFI (Ireland); INFN (Italy); MSIT and NRF (Republic of Korea); MES (Latvia); LMTLT (Lithuania); MOE and UM (Malaysia); BUAP, CINVESTAV, CONACYT, LNS, SEP, and UASLP-FAI (Mexico); MOS (Montenegro); MBIE (New Zealand); PAEC (Pakistan); MSHE, NSC, and NAWA (Poland); FCT (Portugal); MESTD (Serbia); MICIU/AEI and PCTI (Spain); MOSTR (Sri Lanka); Swiss Funding Agencies (Switzerland); MST (Taipei); MHESI (Thailand); TUBITAK and TENMAK (T\"{u}rkiye); NASU (Ukraine); STFC (United Kingdom); DOE and NSF (USA).

\begin{sloppypar}
\setlength\emergencystretch{\hsize}
\hyphenation{Rachada-pisek} Individuals have received support from the Marie-Curie programme and the European Research Council and Horizon 2020 Grant, contract Nos.\ 675440, 724704, 752730, 758316, 765710, 824093, 101115353, 101002207, 101001205, and COST Action CA16108 (European Union); the Leventis Foundation; the Alfred P.\ Sloan Foundation; the Alexander von Humboldt Foundation; the Science Committee, project no. 22rl-037 (Armenia); the Fonds pour la Formation \`a la Recherche dans l'Industrie et dans l'Agriculture (FRIA) and Fonds voor Wetenschappelijk Onderzoek contract No. 1228724N (Belgium); the Beijing Municipal Science \& Technology Commission, No. Z191100007219010, the Fundamental Research Funds for the Central Universities, the Ministry of Science and Technology of China under Grant No. 2023YFA1605804, the Natural Science Foundation of China under Grant No. 12535004, and USTC Research Funds of the Double First-Class Initiative No.\ YD2030002017 (China); the Ministry of Education, Youth and Sports (MEYS) of the Czech Republic; the Shota Rustaveli National Science Foundation (Georgia); the Deutsche Forschungsgemeinschaft (DFG), among others, under Germany's Excellence Strategy -- EXC 2121 ``Quantum Universe" -- 390833306, and under project number 400140256 - GRK2497; the Hellenic Foundation for Research and Innovation (HFRI), Project Number 2288 (Greece); the Hungarian Academy of Sciences, the New National Excellence Program - \'UNKP, the NKFIH research grants K 131991, K 138136, K 143460, K 143477, K 147557, K 146913, K 146914, K 147048, TKP2021-NKTA-64, and 2025-1.1.5-NEMZ\_KI-2025-00004, and MATE KKP and KKPCs Research Excellence and Flagship Research Groups grants (Hungary); the Council of Science and Industrial Research, India; ICSC -- National Research Center for High Performance Computing, Big Data and Quantum Computing, FAIR -- Future Artificial Intelligence Research, and CUP I53D23001070006 (Mission 4 Component 1), funded by the NextGenerationEU program, the Italian Ministry of University and Research (MUR) under Bando PRIN 2022 -- CUP I53C24002390006, PRIN PRIMULA 2022RBYK7T (Italy); the Latvian Council of Science; the Ministry of Science and Higher Education, project no. 2022/WK/14, and the National Science Center, contracts Opus 2021/41/B/ST2/01369, 2021/43/B/ST2/01552, 2023/49/B/ST2/03273, and the NAWA contract BPN/PPO/2021/1/00011 (Poland); the Funda\c{c}\~ao para a Ci\^encia e a Tecnologia (Portugal); the National Priorities Research Program by Qatar National Research Fund;  MICIU/AEI/10.13039/501100011033, ERDF/EU, ``European Union NextGenerationEU/PRTR", projects PID2022-142604OB-C21, PID2022-139519OB-C21, PID2023-147706NB-I00, PID2023-148896NB-I00, PID2023-146983NB-I00, PID2023-147115NB-I00, PID2023-148418NB-C41, PID2023-148418NB-C42, PID2023-148418NB-C43, PID2023-148418NB-C44, PID2024-158190NB-C22, RYC2021-033305-I, RYC2024-048719-I, CNS2023-144781, CNS2024-154769 and Plan de Ciencia, Tecnolog{\'i}a e Innovaci{\'o}n de Asturias, Spain; the Chulalongkorn Academic into Its 2nd Century Project Advancement Project, the National Science, Research and Innovation Fund program IND\_FF\_68\_369\_2300\_097, and the Program Management Unit for Human Resources \& Institutional Development, Research and Innovation, grant B39G680009 (Thailand); the Eric \& Wendy Schmidt Fund for Strategic Innovation through the CERN Next Generation Triggers project under grant agreement number SIF-2023-004; the Kavli Foundation; the Nvidia Corporation; the SuperMicro Corporation; the Welch Foundation, contract C-1845; and the Weston Havens Foundation (USA).
\end{sloppypar}
\end{acknowledgments}\section*{Data availability} Release and preservation of data used by the CMS Collaboration as the basis for publications is guided by the  \href{https://doi.org/10.7483/OPENDATA.CMS.1BNU.8V1W}{CMS data preservation, re-use and open access policy}.
\bibliography{auto_generated} 

\cleardoublepage \appendix\section{The CMS Collaboration \label{app:collab}}\begin{sloppypar}\hyphenpenalty=5000\widowpenalty=500\clubpenalty=5000\cmsinstitute{Yerevan Physics Institute, Yerevan, Armenia}
{\tolerance=6000
A.~Gevorgyan\cmsorcid{0000-0003-2751-9489}, A.~Hayrapetyan, V.~Makarenko\cmsorcid{0000-0002-8406-8605}, A.~Tumasyan\cmsAuthorMark{1}\cmsorcid{0009-0000-0684-6742}
\par}
\cmsinstitute{Marietta Blau Institute for Particle Physics, Vienna, Austria}
{\tolerance=6000
W.~Adam\cmsorcid{0000-0001-9099-4341}, L.~Benato\cmsorcid{0000-0001-5135-7489}, T.~Bergauer\cmsorcid{0000-0002-5786-0293}, M.~Dragicevic\cmsorcid{0000-0003-1967-6783}, S.~Gundacker\cmsorcid{0000-0003-2087-3266}, A.K.~Guven\cmsorcid{0009-0004-5670-5138}, P.S.~Hussain\cmsorcid{0000-0002-4825-5278}, M.~Jeitler\cmsAuthorMark{2}\cmsorcid{0000-0002-5141-9560}, N.~Krammer\cmsorcid{0000-0002-0548-0985}, A.~Li\cmsorcid{0000-0002-4547-116X}, D.~Liko\cmsorcid{0000-0002-3380-473X}, M.~Matthewman, J.~Schieck\cmsAuthorMark{2}\cmsorcid{0000-0002-1058-8093}, R.~Sch\"{o}fbeck\cmsAuthorMark{2}\cmsorcid{0000-0002-2332-8784}, M.~Shooshtari\cmsorcid{0009-0004-8882-4887}, M.~Sonawane\cmsorcid{0000-0003-0510-7010}, N.~Van~Den~Bossche\cmsorcid{0000-0003-2973-4991}, W.~Waltenberger\cmsorcid{0000-0002-6215-7228}, C.E.~Wulz\cmsAuthorMark{2}\cmsorcid{0000-0001-9226-5812}
% ...
\par}
$^{1}$Also at Yerevan State University, Yerevan, Armenia\\
$^{2}$Also at Technische Universit\"{a}t Wien, Vienna, Austria\\
% ...
\end{sloppypar}
\end{document}